\documentclass[prd,aps,nofootinbib,eqsecnum]{revtex4}%,%onecolumn,%groupedaddress]{revtex4}
\usepackage{epsfig}
\usepackage{amsfonts}
\usepackage{rotating}
\usepackage{sparticles}
\usepackage{dcolumn}
\usepackage{bm}
\usepackage{hyperref}
\hypersetup{
    bookmarks=true,         % show bookmarks bar?
    unicode=false,          % non-Latin characters in Acrobat’s bookmarks
    pdftoolbar=true,        % show Acrobat’s toolbar?
    pdfmenubar=true,        % show Acrobat’s menu?
    pdffitwindow=false,     % window fit to page when opened
    pdfstartview={FitH},    % fits the width of the page to the window
    pdftitle={My title},    % title
    pdfauthor={Author},     % author
    pdfsubject={Subject},   % subject of the document
    pdfcreator={Creator},   % creator of the document
    pdfproducer={Producer}, % producer of the document
    pdfkeywords={keyword1} {key2} {key3}, % list of keywords
    pdfnewwindow=true,      % links in new window
    colorlinks=true,       % false: boxed links; true: colored links
    linkcolor=red,          % color of internal links
    citecolor=cyan,        % color of links to bibliography
    filecolor=magenta,      % color of file links
    urlcolor=green,           % color of external links
    linktocpage=true
}
\usepackage{amsmath,amssymb,amsthm,graphicx,latexsym}
\usepackage{subfigure}
\usepackage{pstricks}
\usepackage{color}

\usepackage{mathrsfs}

\newcommand{\be}{\begin{equation}}
\newcommand{\ee}{\end{equation}}
\newcommand{\bea}{\begin{eqnarray}}
\newcommand{\eea}{\end{eqnarray}}
\newcommand{\bml}{\begin{subequations}}
\newcommand{\eml}{\end{subequations}}
\newcommand{\bfig}{\begin{figure}}
\newcommand{\efig}{\end{figure}}

\newcommand{\slashed}{\hspace{-1.1ex}/}

\begin{document}

\title{%Sub-Planckian vs Super-Planckian inflation: \\
\LARGE Can Effective Field Theory of inflation generate large tensor-to-scalar ratio within Randall Sundrum single braneworld? }
\author{Sayantan Choudhury$^{}$\footnote{Electronic address: {sayantan@theory.tifr.res.in, sayanphysicsisi@gmail.com}} ${}^{}$
}
\affiliation{Department of Theoretical Physics, Tata Institute of Fundamental Research, Colaba, Mumbai - 400005, India
\footnote{Presently working as a Visiting (Post-Doctoral) fellow at DTP, TIFR, Mumbai.} ${}^{}$}

%\vspace{5ex}
%\date{\today}
\begin{abstract}
In this paper my prime objective is to explain the generation of large tensor-to-scalar ratio from the single field sub-Planckian inflationary paradigm within Randall Sundrum (RS)
 single braneworld scenario
in a model independent fashion.
By explicit computation I have shown that the effective field theory prescription of brane inflation within RS single brane setup is consistent with sub-Planckian 
 excursion of the inflaton field, which will further generate
large value of tensor-to-scalar ratio, provided the energy density for inflaton degrees of freedom is high enough 
compared to the brane tension in high energy regime. Finally, I have mentioned the stringent
 theoretical constraint on positive brane tension, cut-off of the quantum gravity
scale and bulk cosmological constant to get sub-Planckian field excursion along with large tensor-to-scalar ratio
as recently observed by BICEP2 or at least generates the tensor-to-scalar ratio consistent with the upper bound of Planck (2013 and 2015) data and Planck+BICEP2+Keck Array joint constraint.
\end{abstract}

%\pacs{98.80.-k ; 98.80.Cq ; 04.50.-h}

\maketitle
\tableofcontents
\section{Introduction}
It is a very good-old assumption from superstring theory \cite{Green1,Green2,Polchinski2} that we are living 
in 11 dimensions and different string field theoretic setups are
connected with each other via stringy duality conditions. Among varieties of string theories, 
the 10-dimensional $E_8 \otimes E_8$ heterotic string theory
is a strong candidate for our real world 
as the theory may contain the standard model of particle physics and is related to an 11-dimensional theory written on the orbifold ${\bf R}^{10}
\otimes {\bf S}^1/Z_2$. Within this field theoretic setup, the standard model particle species are confined to the
4-dimensional space-time which is the sub-manifold of ${\bf R}^{4}
\otimes {\bf S}^1/Z_2$. On the contrary, the graviton degrees of freedom propagate
in the total space-time. 
In a most simplified situation, one can think about a 5-dimensional problem where the matter
fields are confined to the 4-dimensional spacetime while gravity 
acts in 5 dimensional bulk spacetime \cite{Maartens:2010ar,Brax:2004xh}.
Amongst very successful propositions for extra dimensional models, Randall \& Sundrum (RS) one brane \cite{Randall:1999vf} and two brane \cite{Randall:1999ee}
models are very famous theoretical prescription in which our observable 
universe in embedded on 3-brane which is 
exactly identical to a domain wall in the context of 5-dimensional anti-de Sitter (${\bf AdS_{5}}$)
space-time. Various cosmo-phenomenological consequences along with inflation have been studied
 from RS setup in
 refs.~\cite{Choudhury:2011sq,Choudhury:2011rz,Choudhury:2012ib,Choudhury:2013yg,Choudhury:2013aqa,Choudhury:2013eoa,Choudhury:2014hna,Das:2013lqa,Banerjee:2014wea,Minamitsuji:2005xs,Himemoto:2000nd,Mukhopadhyaya:2002jn,Ghosh:2008vc,Koley:2005nv}.

The primordial inflation has two key predictions - creating the scalar density perturbations
 and the tensor perturbations during the accelerated phase of expansion \cite{Mukhanov:1981xt,Mukhanov:1990me}.
Very recently, BICEP2~\footnote{BICEP2 result was quite recently put
into question by several works \cite{Liu:2014mpa,Mortonson:2014bja,Flauger:2014qra,Adam:2014gaa}. Also accounting for the
contribution of foreground dust will shift the value of tensor-to-scalar ratio $r$ downward by an amount which will be better
constrained by the joint analysis performed by Planck and BICEP2/Keck Array team \cite{Ade:2015tva}. The final result is expressed as a likelihood curve for r, and yields an upper limit $r_{0.05}<0.12$ at $2\sigma$ confidence.
 Marginalizing over dust and $r$, lensing B-modes are detected at $7\sigma$ significance. Very recently in \cite{Ade:2015lrj} the Planck 
team also fixed the upper bound on the tensor-to-scalar ratio is $r_{0.002} < 0.11$ at $2\sigma$ C.L. and perfectly consistent with the joint analysis performed by Planck and BICEP2/Keck Array team.} \cite{Ade:2014xna} team reported the detection of the primordial tensor perturbations through
the B-mode polarization as:~\be r=0.20^{+0.07}_{-0.05}~ (within~2\sigma~C.L.),\ee 
where $r$ is the tensor-scalar ratio. Explaining this large tensor-to-scalar ratio is a challenging issue for particle cosmologist
because of the Lyth bound \cite{Lyth:1996im}, one would expect a super-Planckian excursion
~\footnote{Field excursion of the inflation filed is defined as: $\Delta \phi=\phi_{cmb}-\phi_{e}$, where $\phi_{cmb}$ represent the field value of the inflaton 
at the momentum scale $k$ which satisfies the equality, $k=aH=-\eta^{-1}\approx k_{*}$, where $(a,~H,~\eta)$ represent the scale factor 
, Hubble parameter, the conformal time and pivot momentum scale respectively. Also $\phi_{e}$ is the field value of the inflaton defined at the end of inflation. Here the super-Planckian excursion is described by, $|\Delta\phi|>M_p$, which is applicable for
large filed models of inflation \cite{Baumann:2014nda,Baumann:2009ds,Linde:2014nna,Kallosh:2014xwa,Kallosh:2014rga}
and sub-Planckian excursion is characterized by, $|\Delta\phi|<M_p$, which hold good in case of small field models of inflation \cite{Lyth:1998xn,Mazumdar:2010sa,Martin:2014vha,Martin:2013nzq}.} of the inflaton field in
order to generate large tensor-to-scalar ratio. It is important to mention here that super-Planckian field excursion computed from the inflationary paradigm is necessarily required to embed the setup with
effective field theory description~\footnote{In case of super-Planckian field excursion it is necessarily required to introduce the higher order quantum 
corrections including the 
effect of higher derivative interactions appearing through the local modifications to GR plays significant role in this context \cite{Choudhury:2013yg}. 
For an example, within 4D Effective Field Theory picture incorporating the local corrections in GR one can write the action as, $$S_{local}=\int d^{4}x
\sqrt{-{}^{(4)}\!g}\left[
\sum^{\infty}_{n=1}{\bf a}_{n}{}^{(4)}\!R^{n}+\sum^{\infty}_{m=1}{\bf b}_{m}\left({}^{(4)}\!R_{\mu\nu}{}^{(4)}\!R^{\mu\nu}\right)^{m}
+\sum^{\infty}_{p=1}{\bf c}_{p}\left({}^{(4)}\!R_{\alpha\beta\delta\eta}{}^{(4)}\!R^{\alpha\beta\delta\eta}\right)^{p}\right].$$   
In this case the appropriate 
choice of the co-efficients ${\bf a_{n}}, {\bf b_{m}}, {\bf c_{p}}$ of the correction factors would
modify the UV behaviour of gravity.
But such local modification of the renormalizable version of GR typically contain
debris like massive ghosts which cannot be regularized or avoided using any field theoretic prescriptions.
If the quantum correction to the usual classical theory of gravity represented via Einstein-Hilbert term is dominated by
 higher derivative nonlocal corrections \cite{Chialva:2014rla,Biswas:2011ar,Biswas:2013cha} then one can avoid such ghost degrees of freedom,
 as the role of these corrections are significant in super-Planckian (or trans-Planckian) scale to make the theory UV complete \cite{Chialva:2014rla}. 
For an example, within 4D Effective Field Theory picture incorporating the non-local corrections in the gravity sector one can write the action as \cite{Biswas:2011ar}:
 \be\begin{array}{lll}
\displaystyle S_{non-local}=
\int d^{4}x\sqrt{-{}^{(4)}\!g}\left[R F_1(\Box)R+R_{\mu\nu} F_2(\Box)R^{\mu\nu}   
+  R_{\mu\nu\alpha\beta} F_{3}(\Box)R^{\mu\nu\alpha\beta}
+ R F_4(\Box)\nabla_{\mu}\nabla_{\nu}\nabla_{\gamma}\nabla_{\eta}R^{\mu\nu\gamma\eta} \right.\\ \left. \displaystyle~~~~~~~~~~~~~~~~~~~~~~~~~~~~~~~~~~~~~~ + 
R_{\mu}^{\nu_1\rho_1\alpha_1} F_{5} (\Box)\nabla_{\rho_1}\nabla_{\alpha_1}\nabla_{\nu_1}\nabla_{\nu}\nabla_{\rho}\nabla_{\alpha}R^{\mu\nu\rho\alpha}
+
R^{\mu_1\nu_1\rho_1\alpha_1} F_{6}(\Box)\nabla_{\rho_1}\nabla_{\alpha_1}
\nabla_{\nu_1}\nabla_{\mu_1}\nabla_{\mu}\nabla_{\nu}\nabla_{\rho}\nabla_{\alpha}R^{\mu\nu\rho\alpha}\right]\nonumber\end{array}\ee
where $F_{i}(\Box)\forall i$ are analytic entire functions containing higher derivatives up to infinite order, where
$\Box=g^{\mu\nu}\nabla_{\mu}\nabla_{\nu}$ is the 4D d'Alembertian operator.
On the other hand in the matter sector incorporating the effects of quantum correction through the interaction between heavy and light (inflaton) field sector 
and finally integrating out the heavy degrees of freedom from the 4D Effective Field Theory picture the matter action, which 
admits a systematic expansion within the light inflaton sector can be written as \cite{Baumann:2014nda,Assassi:2013gxa}:
\be\begin{array}{llll}\label{mod2a}
    \displaystyle S_{matter}[\phi,\Psi]=\int d^{4}x \sqrt{-^{(4)\!}g}
  \left[{\cal L}_{inf}[\phi]+{\cal L}_{heavy}[\Psi]+{\cal L}_{int}[\phi,\Psi]\right]~~\underrightarrow{Remove~~\Psi~}~~e^{i S_{matter}[\phi]}=\int [{\cal D}\Psi] e^{iS_{matter}[\phi,\Psi]}\\
\displaystyle S_{matter}[\phi]=\int d^{4}x \sqrt{-^{(4)\!}g}
  \left[{\cal L}_{inf}[\phi]+\sum_{\alpha}J_{\alpha}(g)\frac{{\cal O}_{\alpha}[\phi]}{M^{\Delta_{\alpha}-4}_{p}}\right]\nonumber
   \end{array}\ee
where $J_{\alpha}(g)$ are dimensionless Wilson coefficients that depend on the couplings g of the UV theory, and ${\cal O}_{\alpha}[\phi]$ are
local operators of dimension $\Delta_{\alpha}$.  This procedure typically generates all possible effective operators ${\cal O}_{\alpha}[\phi]$ consistent with
the symmetries of the UV theory. Also ${\cal L}_{inf}[\phi]$ and ${\cal L}_{heavy}[\Psi]$ describe the part of total Lagrangian density ${\cal L}$ involving only the light and heavy fields, 
and ${\cal L}_{int}[\phi,\Psi]$ includes all
possible interactions involving both sets of fields within Effective Field Theory prescription. After removal of heavy degrees of freedom the effective action is splitted into a renormalizable part:
$${\cal L}_{inf}[\phi]=\frac{g^{\mu\nu}}{2}(\partial_{\mu}\phi)(\partial_{\nu}\phi)-V_{ren}(\phi)$$
 and a sum of non-renormalizable
corrections appearing through the operators ${\cal O}_{\alpha}[\phi]$. Such operators of dimensions less than four 
are called ``relevant operators''. They dominate in the IR and become small in the UV. In 4D Effective Field Theory the operators
of dimensions greater than four are called irrelevant operators. These operators become small in
the IR regime, but dominate in the UV end.
However such corrections are extremely hard to compute and at the same 
time the theoretical origin of all such corrections is not at all clear till now as it completely belongs to the hidden sector of the theory \cite{Assassi:2013gxa}.
One of the possibilities of the origin of such hidden sector heavy field is higher dimensional Superstring Theory or its low energy supergravity version.
 Such a higher dimension setups dimensionally reduced to the 4D Effective Field Theory version via various compactifications. In such a case the corrections arising from graviton loops will always
be weighted by the UV cut-off scale $\Lambda_{UV}$ which is fixed at Planck scale $M_p$, while those coming from heavy sector fields will
be suppressed by the background scale of heavy physics relevant for those fields $M_s$, where
$M_{s}<\Lambda_{UV}\approx M_p$. Present observational status suggests that the scale of such hidden scale is constrained around the GUT scale ($10^{16}$~GeV)
 \cite{Choudhury:2014kma,Choudhury:2014wsa}. In this connection Randall-Sundrum (RS) model is one of possible remedies to solve the trans-Planckian problem
 of field excursion as the 5D cut-off scale of such theory (see section {\bf II} for details) is one order smaller than the 4D cut-off scale of the Effective Field Theory i.e. the 
Planck scale $M_{p}$ to explain the latest ATLAS bound on the lightest
graviton mass
and the Higgs mass within the estimated$\sim$125 GeV against large
radiative correction upto the cut-off of the Model \cite{Das:2013lqa} in the phenomenological ground. In this work using model independent semi-analytical analysis within inflationary 
setup we have explicitly shown that 5D cut-off $M_5$ of RS 
model is also one order smaller than the 4D cut-off scale $M_p$ (see section {\bf III} for details). This also suggests that within RS setup  
the higher order quantum corrections appearing in the gravity as well in the matter sector of the theory is very small in the 4D Effective Field Theory version. 
During our analysis we have further taking an ansatz where the non-renormalizable 4D Planck scale suppressed effective operators only modify the effective potential.
Consequently with the renormalizable part of the potential $V_{ren}$ such corrections will add and finally give rise to the total potential $V(\phi)$ as stated in 
Eq~(\ref{rt10a}). 
}. At present 
it is a very challenging task for the theoretical physicists to propose a new mechanism or technique through which it is possible to 
accommodate sub-Planckian inflation to generate large tensor-to-scalar ratio. The first possibility of addressing this issue is to incorporate the features of 
spectral tilt, running and running of the running by modifying the scale invariant power spectrum. Obviously, the current data can also be explained by the sub-Planckian excursion of the
inflaton field in the context of single field inflation as discussed in \cite{Choudhury:2013iaa,Choudhury:2014kma,Choudhury:2014wsa,Choudhury:2013woa,Choudhury:2014hua},
 where in these class of models sufficient amount of running and running of the running in tensor-to-scalar ratio
has been taken care of. A small class of potentials inspired from particle physics phenomenology i.e. high scale models of inflation in the context of 
MSSM, MSSM$\otimes U(1)_{B-L}$ etc \cite{Choudhury:2013jya,Choudhury:2014sxa,Choudhury:2014uxa,Choudhury:2011jt} will serve this purpose. 
 The next possibility is modified gravity or beyond General Relativistic (GR) framework through which it is possible to address this crucial issue within single field
 inflationary scenario where the effective field theory description holds perfectly. The prime motivation of this work to show explicitly how one can address this
 issue in beyond GR prescription. In this work I investigate the possibility for RS single brane setup in which one can generate
 large tensor-to-scalar ratio along with sub-Planckian field excursion from 
a large class of models of inflation within effective field
 theory prescription 
\cite{Baumann:2014nda,Baumann:2009ds,Cheung:2007st,Weinberg:2008hq,Tsujikawa:2014mba,Senatore:2010wk,LopezNacir:2011kk,Agarwal:2013rva,Assassi:2013gxa,
Baumann:2011nm,Creminelli:2013xfa,Khosravi:2012qg},
and within this setup it is feasible to describe a system through the lowest dimension
operators compatible with the underlying symmetries~\footnote{Assisted 
inflation \cite{Liddle:1998jc,Copeland:1999cs,Kanti:1999ie,Kanti:1999vt,Mazumdar:2001mm,Green:1999vv,Malik:1998gy}
 and N-flation \cite{Dimopoulos:2005ac,Cicoli:2014sva,Easther:2005zr} within multi-field inflationary description, asymptotically free gravity \cite{Chialva:2014rla,Biswas:2011ar,Biswas:2013cha,Stelle:1976gc,Stelle:1977ry,Nunez:2004ts},
shift symmetry \cite{Brax:2005jv,Choudhury:2013zna} are the various possibilities in which it is possible to achieve 
sub-Planckian field excursion along with large tensor-to-scalar ratio and finally the trans-Planckian field excursion issue can be resolved within Effective Field Theory 
prescription.}.

In this paper, I derive the direct connection between field excursion and tensor-so-scalar ratio in the context of effective theory inflation 
within Randall-Sundrum (RS) braneworld scenario
in a model independent fashion. For clarity in the present context the bulk space-time is assumed to have 
5 dimensions. By explicit computation I have shown that the effective field theory of brane inflation within RS setup is consistent with sub-Planckian VEV
 and field excursion, which will further generate
large value of tensor-to-scalar ratio when the energy density for inflaton degrees of freedom is high enough as
compared to the visible and hidden brane tensions in high energy regime. Last but not the least, I have mentioned the stringent
 constraint condition on positive brane tension as well as on the cut-off of the quantum gravity
scale to get sub-Planckian field excursion along with large tensor-to-scalar ratio.

\section{Brane inflation within Radall-Sundrum single brane setup}

Let me start the discussion with a very brief introduction to RS single brane setup.
 The RS single brane setup and its generalized version from a Minkowski brane to a Friedmann-
Robertson-Walker (FRW) brane were derived as solutions
in specific choice of coordinates of the 5D Einstein equations in the bulk, along with the junction conditions which are applied at
the ${\bf Z}_{2}$ -symmetric single brane. A broader perspective, with non-compact dimensions, 
can be obtained via the well known covariant Shiromizu-Maeda-Sasaki approach \cite{Shiromizu:1999wj}, in
which the brane and bulk metrics take its generalized structure. The key point is to use the Gauss-Codazzi
equations to project the 5D bulk curvature along the brane using the covariant formalism. 
Here I start with the well known 5D Rundall Sundrum (RS) single brane model action given by \cite{Randall:1999vf}:
\begin{widetext}
\begin{eqnarray}
 S_{RS}&=& \int d^{5}x\sqrt{-{}^{(5)}\!g}\left[\frac{M^{3}_{5}}{2}~{}^{(5)}\!R-2\Lambda_{5}+{\cal L}_{bulk}+\left({\cal L}_{brane}-\sigma\right)\delta(y)\right],
\end{eqnarray}
\end{widetext}
where the extra dimension ``y'' is non-compact for which the covariant formalism is applicable. Here $M_5$ be the 5D quantum gravity cut-off scale, $\Lambda_5$ be the 5D bulk cosmological constant, ${\cal L}_{bulk}$ be the
bulk field Lagrangian density, ${\cal L}_{brane}$ signifies the Lagrangian density for the brane field contents. 
It is important to mention that the the scalar inflaton degrees of 
freedom is embedded on the 3 brane which has a positive brane tension $\sigma$ and it is localized at the position of orbifold point $y=0$ in case of single brane.
The 5D field
equations in the bulk, including explicitly the contribution of the RS single brane is given by \cite{Shiromizu:1999wj,Maartens:2010ar}:

\begin{equation}
  ^{(5)}\!G_{AB}=\frac{1}{M_5^3}\left[-\Lambda_5 \, {}^{(5)}\!g_{AB}+
  {}^{(5)}T_{AB}+ T_{\mu\nu}^\mathrm{brane}\delta^{\mu}_{A}\delta^{\nu}_{B}\delta(y)\right]
  \label{febfg}
\end{equation}
where $^{(5)}T_{AB}$ characterizes any 5D energy-momentum tensor of the
gravitational sector within bulk specetime. On the other hand the total energy-momentum tensor on the brane
is given by:
  $T_{\mu\nu}^\mathrm{brane} =T_{\mu\nu}-\sigma g_{\mu\nu},$
where $T_{\mu\nu}$ is the energy-momentum tensor of particles and
fields confined to the single brane.
 Further applying the well known Israel--Darmois junction conditions at the brane \cite{Shiromizu:1999wj,Maartens:2010ar} 
finally one can arrive at the $4$-dimensional Einstein induced field equations on the single brane given by \cite{Shiromizu:1999wj,Maartens:2010ar,Brax:2004xh}:

\be 
G_{\mu \nu} = - \Lambda_{4}  g_{\mu \nu} + {1 \over M_p^{2}} T_{\mu \nu} 
+ \left({8\pi \over M_{5}^3}\right)^2 {\cal S}_{\mu \nu} - {\cal E}_{\mu \nu}~,
\label{eq:gmunucvfd}
\ee
where $T_{\mu \nu}$ represents the energy-momentum on the single brane, ${\cal S}_{\mu \nu}$ is a 
rank-2 tensor that contains  contributions that are quadratic in the energy momentum tensor
$T_{\mu \nu}$ \cite{Shiromizu:1999wj,Maartens:2010ar}
and ${\cal E}_{\mu \nu}$ characterizes the projection of the 
5-dimensional Weyl tensor on the 3-brane 
and physically equivalent to the non-local contributions to the pressure and energy flux for a perfect fluid
 \cite{Shiromizu:1999wj,Maartens:2010ar,Brax:2004xh}.

In a cosmological framework, where the 3-brane resembles our universe and the
 metric projected onto the brane is an homogeneous and isotropic flat
 Friedmann-Robertson-Walker (FRW) metric, the  Friedmann
 equation becomes \cite{Shiromizu:1999wj,Maartens:2010ar,Brax:2004xh}:
\be
H^2 = {\Lambda_{4} \over 3} +  {\rho \over 3 M_p^2} 
+ \left({4 \pi \over 3 M_5^3}\right)^2 \rho^2 + {\epsilon \over a^4},
\label{eq:H2}
\ee
where $\epsilon$ is an integration constant. The four and five-dimensional
 cosmological constants are related by \cite{Shiromizu:1999wj,Maartens:2010ar,Brax:2004xh}:
\be
\Lambda_{4} = {4 \pi \over M_5^3} \left(\Lambda_5 + {4 \pi \over 3 M_5^3}~
\sigma^2 \right)~~,
\label{eq:Lam}
\ee
where $\sigma$ is the 3-brane tension. Within RS setup the quantum gravity cut-off scale i.e. the 
5D Planck mass and effective 4D Planck mass are connected through the visible brane tension as:
\begin{eqnarray}\label{mass}
M^{3}_{5}=\sqrt{\frac{4\pi\sigma}{3}}M_{p}.
\end{eqnarray}

Assuming that, as required by observations, the 4D cosmological constant 
is negligible $\Lambda_{4}\approx 0$ in the early universe the localized visible brane tension is given by:
\begin{eqnarray}\label{lam}
 \sigma&=&\sqrt{-\frac{3}{4\pi}M^{3}_{5}\Lambda_{5}}=
\sqrt{-24M^{3}_{5}\tilde{\Lambda}_{5}}>0
\end{eqnarray}
where $\tilde{\Lambda}_{5}$ be the scaled 5D bulk cosmological constant defined as:
\be\label{lamc}
\tilde{\Lambda}_{5}= \frac{\Lambda_{5}}{32\pi}<0.
\ee
 Also the last term 
in Eq. (\ref{eq:H2}) rapidly becomes redundant after inflation sets in, 
the Friedmann equation in RS braneworld becomes \cite{Shiromizu:1999wj,Maartens:2010ar,Brax:2004xh}:
\begin{eqnarray}\label{eq1}
 H^{2}&=&\frac{\rho}{3M^{2}_{p}}\left(1+\frac{\rho}{2\sigma}\right)
\end{eqnarray}
where $\sigma$ be the positive brane tension, $\rho$ signifies the energy density of the inflaton field $\phi$ and $M_{p}=2.43\times 10^{18}~{\rm GeV}$
 be the reduced 4D Planck mass. Using Eq~(\ref{lam}) in Eq~(\ref{mass}), the 5D quantum gravity cut-off scale can be expressed in terms of 5D cosmological constant as:
\begin{eqnarray}\label{mass1}
M^{3}_{5}&=&\sqrt[3]{-\frac{4\pi\Lambda_{5}}{3}}M^{4/3}_{p}=\sqrt[3]{-\frac{128\pi^2\tilde{\Lambda}_{5}}{3}}M^{4/3}_{p}.
\end{eqnarray}
In the low energy limit $\rho<<\sigma$ in which standard GR framework can be retrieved. On the other hand, in the high energy regime $\rho>>\sigma$
as the effect of braneworld correction factor is dominant which is my present focus in this paper. Consequently in high energy limit $\rho>>\sigma$,
Eq~(\ref{eq1}) is written using the slow-roll approximation as:
\begin{eqnarray}\label{eq2}
 H^{2}&\approx&\frac{\rho^{2}}{6M^{2}_{p}\sigma}\approx\frac{V^{2}(\phi)}{6M^{2}_{p}\sigma},
\end{eqnarray}
where $V(\phi)$ be the inflaton single field potential which is expanded in a Taylor series around an intermediate field value $\phi_{i}<\phi_0(<M_p)<\phi_{e}$
~\footnote{Here $\phi_{i}$ and $\phi_{e}$ represent the inflaton field value at the starting point of inflation and at the end of inflation.} as:
\begin{widetext}
\begin{eqnarray}\label{rt10a}
V(\phi)&=&V(\phi_0)+V^{\prime}(\phi_0)(\phi-\phi_{0})+\frac{V^{\prime\prime}(\phi_0)}{2}(\phi-\phi_{0})^{2}
+\frac{V^{\prime\prime\prime}(\phi_0)}{6}(\phi-\phi_{0})^{3}\,%\nonumber\\
%&&~~~~~~~~~~~~~~~~~~~~~~~~~~~~~~~~~~~~~~~~~~~~~~~~
+\frac{V^{\prime\prime\prime\prime}(\phi_0)}{24}(\phi-\phi_{0})^{4}
+\cdots\,\nonumber,\\
&=&\sum^{\infty}_{n=0}\frac{V^{'n}(\phi_0)}{n!}(\phi-\phi_0)^{n},
\end{eqnarray}
\end{widetext}
where $V(\phi_0)\ll M_p^4$ denotes the height of the potential, 
and the coefficients: $V^{\prime}(\phi_0) \leq M_p^3,~V^{\prime\prime}(\phi_0)\leq M_p^2,~V^{\prime\prime\prime}(\phi_0)
\leq M_p,~V^{\prime\prime\prime\prime}(\phi_0)\leq {\cal O}(1)$, determine the shape of the potential in terms
 of the model parameters. The {\it prime} denotes the derivative w.r.t. $\phi$. Here as a special case one can consider a situation
where the intermediate field value $\phi_0$ is identified with the VEV of the inflaton field field $\phi$ i.e. \be\langle 0|\phi| 0\rangle=\phi_0,\ee where $|0\rangle $
be the Bunch-Davies vacuum state using which the VEV is computed in curved space-time. In a most simplest case the numerical value of the VEV is computed from the flatness condition:
\be V^{'}(\phi_0)=0\ee provided $V^{''}(\phi_{0})>0$. In a more advanced situation where inflation is driven by {\it saddle point} and {\it inflection point},
one can impose the flatness constraint on the potential as: \be V^{'}(\phi_0)=0=V^{''}(\phi_0)\ee  for {\it saddle point} \cite{Choudhury:2011jt,Allahverdi:2006iq} and 
\be V^{''}(\phi_0)=0\ee for {\it inflection point} \cite{Choudhury:2013jya,Choudhury:2014sxa,Choudhury:2014uxa,Allahverdi:2006we}
~\footnote{The present observational data from Planck and BICEP2 prefers the {\it inflection point} models of inflation compared to the {\it saddle point}, as the predicted value 
for the scalar spectral tilt obtained from {\it saddle point} inflationary models is low.}. Moreover here it is important mention that the inflaton field belongs to the 
the visible sector of RS setup in which effective field theory prescription perfectly holds good. Even for zero VEV of the inflaton, $\langle 0|\phi |0\rangle =\phi_0=0$, Eq~(\ref{rt10a}) also holds good. 
One can further simplify the expression for the potential by applying ${\bf Z_{2}}$ symmetry in the inflaton field as:
\begin{widetext}
\begin{eqnarray}\label{rt10ab}
V(\phi)&=&V_0+\frac{1}{2}m^{2}\phi^{2}
+\lambda\phi^4+\lambda^{'}M^{-2}_{p}\phi^{6}+\lambda^{''}M^{-4}_{p}\phi^{8}+\cdots\,=
\sum^{\infty}_{m=0}{\bf C}_{2m}\phi^{2m}.
\end{eqnarray}
\end{widetext}
where the expansion co-efficients are defined as:
\begin{eqnarray}
 {\bf C}_{0}&=&V_0,\\
{\bf C}_{2}&=&m^{2}=V^{\prime\prime}(0),\\
{\bf C}_{4}&=&\lambda=\frac{V^{\prime\prime\prime\prime}(0)}{4!},\\
{\bf C}_{6}&=&\lambda^{'}=\frac{M^{2}_{p}V^{\prime\prime\prime\prime\prime\prime}(0)}{6!},\\
{\bf C}_{8}&=&\lambda^{''}=\frac{M^{4}_{p}V^{\prime\prime\prime\prime\prime\prime\prime\prime}(0)}{8!}.%,\\
%&&\cdots\cdots\cdots\cdots\cdots\cdots\cdots\cdots\cdots\nonumber
\end{eqnarray}

Within high energy limit the slow-roll parameters in the visible brane can be expressed as \cite{Maartens:2010ar,Choudhury:2011sq,Choudhury:2012ib}:
\begin{eqnarray}
 \epsilon_{b}(\phi)&\approx& \frac{2M^{2}_{p}\sigma (V^{'}(\phi))^{2}}{V^{3}(\phi)},\\
\eta_{b}(\phi)&\approx& \frac{2M^{2}_{p}\sigma V^{''}(\phi)}{V^{2}(\phi)},\\
\xi^{2}_{b}(\phi)&\approx& \frac{4M^{4}_{p}\sigma^{2} V^{'}(\phi)V^{'''}(\phi)}{V^{4}(\phi)},\\
\sigma^{3}_{b}(\phi)&\approx& \frac{8M^{6}_{p}\sigma^{3} (V^{'}(\phi))^{2}V^{''''}(\phi)}{V^{4}(\phi)}.
\end{eqnarray}
and consequently the number of e-foldings can be written as \cite{Maartens:2010ar,Choudhury:2011sq,Choudhury:2012ib}:
\begin{eqnarray}\label{efolda}
\Delta N_{b}=|N_{b}(\phi_{cmb})-N_{b}(\phi_{e})|&\approx&  \frac{1}{2\sigma M^2_p}\int^{\phi_{cmb}}_{\phi_{e}} d\phi\frac{V^{2}(\phi)}{V^{'}(\phi)}
\end{eqnarray}
where $\phi_{e}$ corresponds to the field value at the end of inflation which can be obtained from 
the following equation:
\begin{eqnarray}
 \max_{\phi=\phi_{e}}\left[\epsilon_{b},|\eta_{b}|,|\xi^{2}_{b}|,|\sigma^{3}_{b}|\right]&=&1.
\end{eqnarray}

In terms of the momentum, the number of
e-foldings, ${N}_{b}(k)$,  can be expressed as~\cite{Burgess:2005sb}:
\begin{widetext}
\be\begin{array}{llll}\label{efold}
\displaystyle {N}_{b}(k) \approx  71.21 - \ln \left(\frac{k}{k_{*}}\right)  
+  \frac{1}{4}\ln{\left( \frac{V_{*}}{M^4_{P}}\right) }
+  \frac{1}{4}\ln{\left( \frac{V_{*}}{\rho_{e}}\right) }  
+ \frac{1-3w_{int}}{12(1+w_{int})} 
\ln{\left(\frac{\rho_{rh}}{\rho_{e}} \right)},
\end{array}\ee
\end{widetext}
where $\rho_{e}$ is the energy density at the end of inflation, 
$\rho_{rh}$ is an energy scale during reheating, 
$k_{*}=a_* H_*$ is the present Hubble scale, 
$V_{*}$ corresponds to the potential energy when the relevant modes left the Hubble patch 
during inflation corresponding to the momentum scale $k_{*}$, and $w_{int}$ characterises the effective equation of state 
parameter between the end of inflation, and the energy scale during reheating. 
Within the momentum interval,
$k_{e}<k<k_{cmb}$, the corresponding number of e-foldings is given by, $\Delta {N}_{b}$, as
\begin{widetext}
\begin{eqnarray}\label{intnk}
    \Delta {N}_{b} =|{N}_{b}(k_{e})-{N}_{b}(k_{cmb})|&=&\ln\left(\frac{k_{cmb}}{k_{e}}\right)=
\ln\left(\frac{a_{cmb}}{a_{e}}\right)+\ln\left(\frac{H_{cmb}}{H_{e}}\right)=\ln\left(\frac{a_{cmb}}{a_{e}}\right)+\ln\left(\frac{V_{cmb}}{V_{e}}\right)\,
   \end{eqnarray}
\end{widetext}
where $(a_{cmb},H_{cmb})$ and $(a_{e}H_{e})$
represent the scale factor and the Hubble parameter at the CMB scale and end of inflation.
 One can estimate the contribution of the last term of the right hand side by using Eq~(\ref{rt10a}) as:
 \begin{widetext}\be\begin{array}{lll}\label{vcmb}\displaystyle\left(\frac{V_{cmb}}{V_{e}}\right)=\left[1+\sum^{\infty}_{n=1}
\frac{V^{\prime n}(\phi_0)}{n!V(\phi_0)}(\phi_{cmb}-\phi_{0})^{n}\right]
\left[1+\sum^{\infty}_{j=1}\frac{V^{\prime j}(\phi_0)}{j!V(\phi_0)}(\phi_{e}-\phi_{0})^{j}\right]^{-1},\\
\displaystyle ~~~~~~~~~~~~\approx\left[1+\sum^{\infty}_{n=1}\frac{V^{\prime n}(\phi_0)}{n!V(\phi_0)}(\phi_{cmb}-\phi_{0})^{n}
-\sum^{\infty}_{j=1}\frac{V^{\prime j}(\phi_0)}{j!V(\phi_0)}(\phi_{e}-\phi_{0})^{j}-\sum^{\infty}_{n=1}
\sum^{\infty}_{j=1}\frac{V^{\prime n}(\phi_0)V^{\prime j}(\phi_0)}{n!j!V^2(\phi_0)}(\phi_{cmb}-\phi_{0})^{n}(\phi_{e}-\phi_{0})^{j}\right],\\
\displaystyle ~~~~~~~~~~~~\approx\left[1+W-Q \right],\end{array}\ee
  \end{widetext}
where $W$ and $Q$ represent two series sum given by:
\begin{widetext} 
\begin{eqnarray}
     W&=&\sum^{\infty}_{j=1}\frac{1}{(j-1)!}\left(\frac{\Delta\phi}{M_p}\right)\frac{V^{\prime j}(\phi_0)M^{j}_{p}}{V(\phi_0)}
\left(\frac{\phi_{e}-\phi_{0}}{M_p}\right)^{j-1},\\
Q&=&\sum^{\infty}_{n=1}
\sum^{\infty}_{j=1}\frac{V^{\prime n}(\phi_0)V^{\prime j}(\phi_0)M^{n+j}_{p}}{V^2(\phi_0)}\left\{\frac{1}{n!j!}\left(\frac{\phi_{e}-\phi_{0}}{M_{p}}\right)^{n+j}+\frac{1}{(n-1)!j!}\left(\frac{\Delta\phi}{M_p}\right)
\left(\frac{\phi_{e}-\phi_{0}}{M_{p}}\right)^{n+j-1}\right\}
   \end{eqnarray}
\end{widetext}
where the field excursion is defined as, $\Delta\phi=\phi_{cmb}-\phi_{e}$, where $\phi_{cmb}$ and $\phi_{e}$ signify the 
inflaton field value at the at the last scattering surface (LSS) of CMB or more precisely at the horizon crossing~\footnote{Here horizon crossing stands for the 
physical situation where the corresponding momentum scale satisfies the equality $k=\frac{2\pi}{\lambda_{w}}=aH$, where $\lambda_{w}$ be the associated wavelength of the scalar and tensor modes whose snapshot are 
observed at the LSS of CMB. After crossing the horizon
 all such modes goes to the super-Hubble region in which the momentum scale $k>>aH$ i.e. $\lambda_{w}<<\frac{2\pi}{aH}$,
 which implies the corresponding wavelengths of the scalar snd tensor modes are too small to be detected. On the other hand, before the horizon crossing there will be region in a smooth patch within sub-Hubble region where the corresponding 
momentum scale $k<<aH$ i.e. $\lambda_{w}>>\frac{2\pi}{aH}$, which can be detected via various observational probes.} and at the end of inflation respectively. Now I explicitly show that both of the series sum are convergent in the present context.
To hold the effective field theory prescription one need to satisfy the following sets of
criteria:
\begin{itemize}
\item {\bf (1).} ${\large\left(\frac{\phi_{e}-\phi_{0}}{M_p}\right)\leq 1}$,~~%\\
\item {\bf (2).} $\left(\frac{\Delta\phi}{M_p}\right)\leq 1$,
\item {\bf (3).} $\frac{V^{\prime j}(\phi_0)M^{j}_{p}}{V(\phi_0)}\leq 1\forall j$,
\item {\bf (4).} $\frac{V^{\prime n}(\phi_0)V^{\prime j}(\phi_0)M^{n+j}_{p}}{V^2(\phi_0)}\leq 1\forall (n,j)$.
\end{itemize}
This implies that, both $W<1$ and $Q<1$ are convergent and from Eq~(\ref{vcmb}) we get:
\be
\left(\frac{V_{cmb}}{V_{e}}\right)\approx 1,\ee
which perfectly holds good for zero VEV inflaton case.
Let us investigate the ${\bf Z_{2}}$ symmetric case in which one can write:
\begin{widetext}\be\begin{array}{lll}\label{vcmb}\displaystyle\left(\frac{V_{cmb}}{V_{e}}\right)=\left[1+\sum^{\infty}_{n=1}
\frac{{\bf C}_{2n}}{V_0}\phi^{2n}_{cmb}\right]
\left[1+\sum^{\infty}_{j=1}\frac{{\bf C}_{2j}}{V_0}\phi^{2j}_{e}\right]^{-1},\\
\displaystyle ~~~~~~~~~~~~\approx\left[1+\sum^{\infty}_{n=1}
\frac{{\bf C}_{2n}}{V_0}\phi^{2n}_{cmb}
-\sum^{\infty}_{j=1}\frac{{\bf C}_{2j}}{V_0}\phi^{2j}_{e}-\sum^{\infty}_{n=1}
\sum^{\infty}_{j=1}\frac{{\bf C}_{2n}{\bf C}_{2j}}{V^2_0}\phi^{2n}_{cmb}\phi^{2j}_{e}\right],\\
\displaystyle ~~~~~~~~~~~~\approx\left[1+W_{0}-Q_{0} \right],\end{array}\ee
  \end{widetext}
where $W_{0}$ and $Q_{0}$ represent two series sum given by:
\begin{widetext} 
\begin{eqnarray}
     W_{0}&=&2\sum^{\infty}_{j=1}j\left(\frac{\Delta\phi}{M_p}\right)\frac{{\bf C}_{2j}M^{2j}_{p}}{V_0}
\left(\frac{\phi_{e}}{M_p}\right)^{2j-1},\\
Q_{0}&=&\sum^{\infty}_{n=1}
\sum^{\infty}_{j=1}\frac{{\bf C}_{2n}{\bf C}_{2j}M^{2(n+j)}_{p}}{V^2_0}\left\{\left(\frac{\phi_{e}}{M_{p}}\right)^{2(n+j)}+2n\left(\frac{\Delta\phi}{M_p}\right)
\left(\frac{\phi_{e}}{M_{p}}\right)^{2(n+j)-1}\right\}
   \end{eqnarray}
\end{widetext}
Here also the similar criteria hold good to apply the effective field theory prescription which make the series sum $W_{0}$ and $Q_{0}$ convergent. 
Consequently, for all the physical situations described in this paper Eq~(\ref{intnk}) reduces to:
\begin{eqnarray}\label{intnk1}
     \Delta {N}_{b} &\approx& \ln\left(\frac{k_{cmb}}{k_{e}}\right)=\ln\left(\frac{a_{cmb}}{a_{e}}\right).
   \end{eqnarray}

\section{Field excursion within effective theory description}
In the high energy limit of RS braneworld the tensor-to-scalar ratio satisfies the following consistency condition at the leading order of the effective field theory:
\begin{eqnarray}\label{rk}
 r_{b}(k)&=&\frac{P_{T}(k)}{P_{S}(k)}=24\epsilon_{b}=\frac{48M^{2}_{p}\sigma (V^{'}(\phi))^2}{V^{3}(\phi)}
\end{eqnarray}
where $P_{S}(k)$ and $P_{T}(k)$ are the scalar and tensor power spectrum at any scale $k$. It is important to note that 
the following operator relationship holds good in the high energy limit of RS braneworld:
\begin{eqnarray}\label{id}
 \frac{d}{d\phi}=-\frac{V^{2}}{2\sigma M^{2}_{p}V^{'}}\frac{d}{d\ln k}.
\end{eqnarray}
In Eq~(\ref{rk})
the tensor-to-scalar ratio can be parametrized at any arbitrary momentum scale as:
\begin{widetext}
\be\begin{array}{lll}\label{rk1}
  \displaystyle r_{b}(k)\displaystyle =\left\{\begin{array}{ll}
                    \displaystyle  r_{b}(k_{*}) &
 \mbox{ {\bf for \underline{Case I}}}  \\ 
         \displaystyle  r_{b}(k_{*})\left(\frac{k}{k_{*}}\right)^{n_{T}(k_{*})-n_{S}(k_{*})+1} & \mbox{ {\bf for \underline{Case II}}}\\ 
\displaystyle  r_{b}(k_{*})\left(\frac{k}{k_{*}}\right)^{n_{T}(k_{*})-n_{S}(k_{*})+1+\frac{\alpha_{T}(k_{*})-\alpha_{S}(k_{*})}{2!}\ln\left(\frac{k}{k_{*}}\right)} & \mbox{ {\bf for \underline{Case III}}} \\
\displaystyle  r_{b}(k_{*})\left(\frac{k}{k_{*}}\right)^{n_{T}(k_{*})-n_{S}(k_{*})+1+\frac{\alpha_{T}(k_{*})-\alpha_{S}(k_{*})}{2!}\ln\left(\frac{k}{k_{*}}\right)
+\frac{\kappa_{T}(k_{*})-\kappa_{S}(k_{*})}{3!}\ln^2\left(\frac{k}{k_{*}}\right)} & \mbox{ {\bf for \underline{Case IV}}}.
          \end{array}
\right.
\end{array}\ee
\end{widetext}
where $k_{*}$ be the pivot scale of momentum. In Eq~(\ref{rk1}) the subscript $(T,S)$ signifies the tensor and scalar modes
obtained from cosmological perturbation in RS braneworld. Here $(n_{T},n_{S})$, $(\alpha_{T},\alpha_{S})$ and $(\kappa_{T},\kappa_{S})$ represent the tensor and
scalar spectral tilt, running and running of the running in RS braneworld respectively. See appendix where all these definitions are explicitly given.
Also in Eq~(\ref{rk1}) I mention four possibilities as given by:
\begin{itemize}
                        \item {\bf Case I} stands for a situation where the spectrum is scale invariant,
\item  {\bf Case II} stands for a situation where spectrum follows power law feature 
through the spectral tilt $(n_{S},n_{T})$,
\item {\bf Case III} signifies a situation where the spectrum shows deviation from power low in presence of running of the
 spectral tilt $(\alpha_{S},\alpha_{T})$ along with logarithmic correction in the momentum scale (as appearing in the exponent) and
\item {\bf Case IV} characterizes
 a physical situation in which the spectrum is further modified compared to the {\bf Case III},
by allowing running of the running of spectral tilt $(\kappa_{S},\kappa_{T})$ along with square of the momentum dependent logarithmic correction.
                       \end{itemize}

Further combining Eq~(\ref{rk}) and Eq~(\ref{id}) together and performing the momentum as well as the slow-roll integration I get:
\begin{eqnarray}\label{con2}
 \frac{1}{2}\sqrt{\frac{\sigma}{3}}\left|\int^{k_{cmb}}_{k_{e}}d\ln k ~\sqrt{r_{b}(k)}\right|&=& \frac{1}{M_{p}}\left|\int^{\phi_{cmb}}_{\phi_{e}}d\phi~\sqrt{V(\phi)}\right|.\nonumber\\
\end{eqnarray}

Finally substituting Eq~(\ref{rk3}) and Eq~(\ref{rk8}) on Eq~(\ref{con2}) I get: 
\begin{widetext}
\be\begin{array}{lll}\label{rk9}
 \displaystyle\left|\frac{\Delta\phi}{M_p}\right|\displaystyle =\frac{1}{2}\sqrt{\frac{\sigma}{3{\cal V}_{inf}}}\times\left\{\begin{array}{ll}
                    \displaystyle  \sqrt{r_{b}(k_*)}|\Delta {N}_{b}| &
 \mbox{\small {\bf for \underline{Case I}}}  \\ 
         \displaystyle  \frac{2\sqrt{r_{b}(k_*)}}{n_{T}(k_{*})-n_{S}(k_{*})+1}\left|1-e^{-\Delta {N}_{b}\left(\frac{n_{T}(k_{*})-n_{S}(k_{*})+1}{2}\right)}\right| & \mbox{\small {\bf for \underline{Case II}}}\\
\displaystyle  \sqrt{r_{b}(k_*)}e^{-\frac{(n_{T}(k_{*})-n_{S}(k_{*})+1)^2}{2(\alpha_{T}(k_{*})-\alpha_{S}(k_{*}))}}\sqrt{\frac{2\pi}{(\alpha_{T}(k_{*})-\alpha_{S}(k_{*}))}}
\\ \displaystyle \left|{\rm erfi}\left(\frac{n_{T}(k_{*})-n_{S}(k_{*})+1}{\sqrt{
2(\alpha_{T}(k_{*})-\alpha_{S}(k_{*}))}}\right)
\right.\\ \left. \displaystyle ~~~~~~
-{\rm erfi}\left(\frac{n_{T}(k_{*})-n_{S}(k_{*})+1}{\sqrt{2(\alpha_{T}(k_{*})-\alpha_{S}(k_{*}))}}-\sqrt{\frac{(\alpha_{T}(k_{*})-\alpha_{S}(k_{*}))}{8}}
\Delta {N}_{b}\right)\right| & \mbox{\small {\bf for \underline{Case III}}}\\ 
\displaystyle  \sqrt{r_{b}(k_{*})}\left|\left(\frac{3}{2}-\frac{n_{T}(k_{*})-n_{S}(k_{*})}{2}+\frac{\alpha_{T}(k_{*})-\alpha_{S}(k_{*})}{8}\right.\right.\\ \left.\left. 
\displaystyle ~~~~~\displaystyle -\frac{\kappa_{T}(k_{*})-\kappa_{S}(k_{*})}{24}\right)\displaystyle \left\{1-e^{-\Delta {N}_{b}}\right\}-
\left(\frac{1}{2}-\frac{n_{T}(k_{*})-n_{S}(k_{*})}{2}\right.\right.\\ \left.\left.\displaystyle \displaystyle+\frac{\alpha_{T}(k_{*})-\alpha_{S}(k_{*})}{8} 
 -\frac{\kappa_{T}(k_{*})-\kappa_{S}(k_{*})}{24}\right)\displaystyle 
\Delta {N}_{b}e^{-\Delta {N}_{b}}\right.\\ \left.\displaystyle -\left(\frac{\kappa_{T}(k_{*})
-\kappa_{S}(k_{*})}{48}-\frac{\alpha_{T}(k_{*})-\alpha_{S}(k_{*})}{16}\right)\displaystyle
(\Delta {N}_{b})^{2}e^{-\Delta {N}_{b}}\right.\\ \left.\displaystyle 
-\frac{\kappa_{T}(k_{*})
-\kappa_{S}(k_{*})}{144}\displaystyle
(\Delta {N}_{b})^{3}e^{-\Delta {N}_{b}}\right| & \mbox{\small {\bf for \underline{Case IV}}}.
          \end{array}
\right.
\end{array}\ee
\end{widetext}
Here all the observables appearing in the left side of Eq~(\ref{rk9}) can also be expressed in terms of the slow-roll parameters in RS single braneworld. See the appendix for details. 
Further using the the limiting results on $\Delta N_{b}$ I get:
\begin{widetext}
\be\begin{array}{lll}\label{rk10}
 \displaystyle\lim_{\Delta {N}_{b}\rightarrow small}\left|\frac{\Delta\phi}{M_p}\right|\displaystyle =\frac{1}{2}\sqrt{\frac{\sigma}{3{\cal V}_{inf}}}\times\left\{\begin{array}{ll}
                   % \displaystyle  \sqrt{r_{b}(k_*)}|\Delta {N}_{b}| &
 %\mbox{\small {\bf for \underline{Case I}}}  \\ \\
         \displaystyle  \sqrt{r_{b}(k_*)}|\Delta {N}_{b}| & \mbox{\small {\bf for \underline{Case II}}}\\ 
\displaystyle  \sqrt{r_{b}(k_*)}|\Delta {N}_{b}|e^{-\frac{(n_{T}(k_{*})-n_{S}(k_{*})+1)^2}{2(\alpha_{T}(k_{*})-\alpha_{S}(k_{*}))}}\ & \mbox{\small {\bf for \underline{Case III}}}\\
\displaystyle  \sqrt{r_{b}(k_*)}|\Delta {N}_{b}|\left|1-\left(\frac{\kappa_{T}(k_{*})
-\kappa_{S}(k_{*})}{48}-\frac{\alpha_{T}(k_{*})-\alpha_{S}(k_{*})}{16}\right)\displaystyle
\Delta {N}_{b}\right.\\ \left.\displaystyle 
~~~~~~~~~~~~~~~~~~~~~~~~~~~~~~~~~~~~~~~~~~-\frac{\kappa_{T}(k_{*})
-\kappa_{S}(k_{*})}{144}\displaystyle
(\Delta {N}_{b})^{2}\right| & \mbox{\small {\bf for \underline{Case IV}}}.
          \end{array}
\right.
\end{array}\ee
\end{widetext}
 Most importantly Eq~(\ref{efolda1}) and Eq~(\ref{efolda2}) 
fix the value of $\Delta N_{b}$ within the desired range demanded by the observational probes. This can be easily done by putting constraint on
the brane tension of the single brane and the Taylor expansion co-efficients of the effective potential within RS setup. Also this makes the analysis consistent 
presented in this paper. Further from Eq~(\ref{efolda1}) and Eq~(\ref{efolda2}) one can write the field excursion for the both the physical situations as:
\be\begin{array}{llll}\label{efolda1a}
\underline{\bf Without~~Z_2}:~~~~~
\displaystyle \left|\frac{2\sigma \Delta N_{b} V^{'}(\phi_0)M_{p}}{V^{2}(\phi_0)}\right|\approx 
\displaystyle
\left|\frac{\Delta\phi}{M_p}\right|\leq 1,\end{array}\ee
\be\begin{array}{llll}\label{efolda2a}
\underline{\bf With~~Z_2}:~~~~~
\displaystyle\left|\frac{4\sigma\phi_e\Delta N_{b}m^2 M_p}{V^{2}_0}\right|\approx\displaystyle \left|\frac{\Delta\phi}{M_{p}}\right|\leq 1.\end{array}\ee
Now using Eq~(\ref{efolda1a}) and Eq~(\ref{efolda2a})
one can express the analytical bound on the positive brane tension $\sigma$ as:
\be\begin{array}{llll}\label{efolda1aa}
\underline{\bf Without~~Z_2}:~~~~~
\displaystyle \sigma\leq 
\displaystyle
\left|\frac{V^{2}(\phi_0)}{2\Delta N_{b} V^{'}(\phi_0)M_p}\right|,\end{array}\ee
\be\begin{array}{llll}\label{efolda2aa}
\underline{\bf With~~Z_2}:~~~~~
\displaystyle\sigma\leq\displaystyle \left|\frac{V^{2}_0}{4\phi_e \Delta N_b m^2 M_p}\right|.\end{array}\ee

Now I will explicitly show the 
details of each of the constraints on $\sigma$ computed from Eq~(\ref{efolda1aa}) and Eq~(\ref{efolda2aa}). To serve this purpose
let me now first write down the Taylor expansion co-efficient of the generic potential
$V(\phi_{*}),V^{\prime}(\phi_{*}),V^{\prime\prime}(\phi_{*}),\cdots$ in terms of the inflationary observables:
\be\begin{array}{llll}\label{aq1}
\displaystyle V(\phi_{*})= \sqrt[3]{2\pi^2 P_{S}(k_{*})r(k_{*})}M^{4/3}_{p}\sigma^{2/3},\\
\displaystyle V^{'}(\phi_{*})= \sqrt{\frac{P_{S}(k_{*})\sigma}{24}}~\pi~r(k_{*})M_{p},\\
\displaystyle V^{''}(\phi_{*})= 2^{-4/3}(P_{S}(k_{*})r(k_{*}))^{2/3}\pi^{4/3}\left(n_{S}(k_{*})-1+\frac{r(k_{*})}{4}\right)M^{2/3}_{p}\sigma^{1/3},\\
\displaystyle V^{'''}(\phi_{*})= 2^{-5/3}(P_{S}(k_{*})r(k_{*}))^{4/3}\pi^{5/3}\left[\frac{r(k_{*})}{3}\left(n_{S}(k_{*})-1+\frac{r(k_{*})}{4}\right)
%\right.\\ \left.~~~~~~~~~~~~~~~~~~~~~~~~~~~~~~~~~~~~~~~~~~~~~~~~~~~~~~~~~~~~\displaystyle
 -18
\left(\frac{r(k_{*})}{24}\right)^{2}-\alpha_{S}(k_{*})\right]M^{1/3}_{p}\sigma^{1/6},\\
\displaystyle V^{''''}(\phi_{*})=\frac{V^{4}(\phi_*)}{8M^{6}_{p}(V^{'}(\phi_*))^2}\left[\frac{\kappa_{S}(k_{*})}{2}-
4\left(\frac{r(k_{*})}{8}\right)^{2}\left(n_{S}(k_{*})-1+\frac{r(k_{*})}{4}\right)\right.\\ \left.\displaystyle 
~~~~~~~~~~~~~~~~~~~~~~~~~~~~~~~~~~~~~~~~~~~~~~~~~~
+96\left(\frac{r(k_{*})}{24}\right)^{3}+\frac{r(k_{*})}{3}\left(n_{S}(k_{*})-1+\frac{r(k_{*})}{4}\right)^{2}
\right.\\ \left.\displaystyle 
~~~~~~~~~~~~~~~~~~~~~~~~~~~~~~~~~~~~~~~~~~~~~~~~~~-\frac{4M^{4}_{p}\sigma^{2}(V^{'}(\phi_*))^2 V^{'''}(\phi_*)}{V^{4}(\phi_*)}\left(n_{S}(k_{*})-1-\frac{r(k_{*})}{12}\right)\right],\\
\cdots\quad\cdots\quad\cdots\quad\cdots\quad\cdots\quad\cdots\quad\cdots\quad\cdots\quad\cdots\quad\cdots\quad\cdots\quad\cdots\quad\cdots\quad\cdots\quad\cdots\quad\cdots 
\quad\cdots\quad\cdots\quad\cdots\quad\cdots\quad\cdots
\end{array}\ee
where I use the fact that inflaton field value at the pivot scale $\phi_{*}\approx \phi_{cmb}$.
Therefore, one can write a matrix equation characterizing the Taylor expansion coefficients at VEV $\phi_{0}$ as:
\begin{equation}\label{mat1}\left(\begin{tabular}{ccccccc}
 $1$ & $\Theta_{*}$ & $\frac{\Theta^{2}_{*}}{2}$& $\frac{\Theta^{3}_{*}}{6}$ & $\frac{\Theta^{4}_{*}}{24}$&$\cdots$&$\cdots$\\
$0$ & $1$ & $\Theta_{*}$& $\frac{\Theta^{2}_{*}}{2}$ & $\frac{\Theta^{3}_{*}}{6}$&$\cdots$&$\cdots$\\
$0$ & 0 & 1& $\Theta_{*}$ & $\frac{\Theta^{2}_{*}}{2}$&$\cdots$&$\cdots$\\
$0$ & $0$ & $0$& $1$ & $\Theta_{*}$&$\cdots$&$\cdots$\\
$0$ & $0$ & $0$& $0$ & $1$&$\cdots$&$\cdots$\\
$\cdots$ & $\cdots$ & $\cdots$& $\cdots$ & $\cdots$&$\cdots$&$\cdots$\\
$\cdots$ & $\cdots$ & $\cdots$& $\cdots$ & $\cdots$&$\cdots$ &$\cdots$\\ 
      \end{tabular}\right)
\left(\begin{tabular}{c}
      $V(\phi_{0})$\\
      $V^{'}(\phi_0)$\\
      $V^{''}(\phi_0)$\\
      $V^{'''}(\phi_0)$\\
      $V^{''''}(\phi_0)$\\
$\cdots$\\
$\cdots$\\
      \end{tabular}\right)
\begin{tabular}{c}
      \\
      =\\
      \\
      \end{tabular}
\left(\begin{tabular}{c}
      $V(\phi_{*})$\\
      $V^{'}(\phi_{*})$\\
      $V^{''}(\phi_{*})$\\
     $V^{'''}(\phi_{*})$\\
      $V^{''''}(\phi_{*})$\\
$\cdots$\\
$\cdots$\\
      \end{tabular}\right)\,,
\end{equation}
where I introduce a new symbol:
\be\Theta_{*}:=(\phi_{*}-\phi_0)=\left(\underbrace{\frac{\phi_{e}-\phi_{0}}{M_p}}_{\leq 1}+\underbrace{\frac{\Delta\phi}{M_p}}_{\leq 1}\right)M_p\leq M_p.\ee
Finally applying the matrix inversion technique I get the following physical solution:
\begin{equation}\label{mat2}
\left(\begin{tabular}{c}
      $V(\phi_{0})$\\
      $V^{'}(\phi_0)$\\
      $V^{''}(\phi_0)$\\
      $V^{'''}(\phi_0)$\\
      $V^{''''}(\phi_0)$\\
      $\cdots$\\
      $\cdots$\\
\end{tabular}\right)
\begin{tabular}{c}
      \\
      =\\
      \\
      \end{tabular}
\left(\begin{tabular}{ccccccc}
 $1$ & $-\Theta_{*}$ & $\frac{\Theta^{2}_{*}}{2}$& $-\frac{\Theta^{3}_{*}}{6}$ & $\frac{\Theta^{4}_{*}}{24}$&$\cdots$&$\cdots$\\
$0$ & $1$ & $-\Theta_{*}$& $\frac{\Theta^{2}_{*}}{2}$ & $-\frac{\Theta^{3}_{*}}{6}$&$\cdots$&$\cdots$\\
$0$ & 0 & 1& $-\Theta_{*}$ & $\frac{\Theta^{2}_{*}}{2}$&$\cdots$&$\cdots$\\
$0$ & $0$ & $0$& $1$ & $-\Theta_{*}$&$\cdots$&$\cdots$\\
$0$ & $0$ & $0$& $0$ & $1$&$\cdots$&$\cdots$\\
$\cdots$&$\cdots$&$\cdots$&$\cdots$&$\cdots$&$\cdots$&$\cdots$\\
$\cdots$&$\cdots$&$\cdots$&$\cdots$&$\cdots$&$\cdots$&$\cdots$\\
      \end{tabular}\right)
\left(\begin{tabular}{c}
      $V(\phi_{*})$\\
      $V^{'}(\phi_{*})$\\
      $V^{''}(\phi_{*})$\\
     $V^{'''}(\phi_{*})$\\
      $V^{''''}(\phi_{*})$\\
      $\cdots$\\
      $\cdots$\\
      \end{tabular}\right).
\end{equation}
As the series converge criteria holds good in the present context, one can write down the following solution in the leading order approximation as: 
\begin{equation}\label{mat3}
\left(\begin{tabular}{c}
      $V(\phi_{0})$\\
      $V^{'}(\phi_0)$\\
      $V^{''}(\phi_0)$\\
      $V^{'''}(\phi_0)$\\
      $V^{''''}(\phi_0)$\\
      $\cdots$\\
      $\cdots$\\
\end{tabular}\right)
\begin{tabular}{c}
      \\
      $\approx$\\
      \\
      \end{tabular}
\left(\begin{tabular}{c}
      $V(\phi_{*})$\\
      $V^{'}(\phi_{*})$\\
      $V^{''}(\phi_{*})$\\
     $V^{'''}(\phi_{*})$\\
      $V^{''''}(\phi_{*})$\\
      $\cdots$\\
      $\cdots$\\
      \end{tabular}\right).
\end{equation}
Now in case of ${\bf Z}_{2}$ symmetric situation with zero VEV one can rewrite the solution of matrix equation as:
\begin{equation}\label{mat4}
\left(\begin{tabular}{c}
      $V_{0}$\\
      %$0$\\
      $m^{2}$\\
      %$0$\\
      $24\lambda$\\
      $\cdots$\\
      $\cdots$\\
$\cdots$\\
$\cdots$\\
\end{tabular}\right)
\begin{tabular}{c}
      \\
      =\\
      \\
      \end{tabular}
\left(\begin{tabular}{ccccccc}
 $1$ & $-\phi_{*}$ & $\frac{\phi^{2}_{*}}{2}$& $-\frac{\phi^{3}_{*}}{6}$ & $\frac{\phi^{4}_{*}}{24}$&$\cdots$&$\cdots$\\
%$0$ & $1$ & $-\phi_{*}$& $\frac{\phi^{2}_{*}}{2}$ & $-\frac{\phi^{3}_{*}}{6}$&$\cdots$&$\cdots$\\
$0$ & 0 & 1& $-\phi_{*}$ & $\frac{\phi^{2}_{*}}{2}$&$\cdots$&$\cdots$\\
%$0$ & $0$ & $0$& $1$ & $-\phi_{*}$&$\cdots$&$\cdots$\\
$0$ & $0$ & $0$& $0$ & $1$&$\cdots$&$\cdots$\\
$\cdots$&$\cdots$&$\cdots$&$\cdots$&$\cdots$&$\cdots$&$\cdots$\\
$\cdots$&$\cdots$&$\cdots$&$\cdots$&$\cdots$&$\cdots$&$\cdots$\\
$\cdots$&$\cdots$&$\cdots$&$\cdots$&$\cdots$&$\cdots$&$\cdots$\\
$\cdots$&$\cdots$&$\cdots$&$\cdots$&$\cdots$&$\cdots$&$\cdots$\\
      \end{tabular}\right)
\left(\begin{tabular}{c}
      $V(\phi_{*})$\\
      %$V^{'}(\phi_{*})$\\
      $V^{''}(\phi_{*})$\\
     %$V^{'''}(\phi_{*})$\\
      $V^{''''}(\phi_{*})$\\
      $\cdots$\\
      $\cdots$\\
$\cdots$\\
$\cdots$\\
      \end{tabular}\right).
\end{equation}
where 
\be\phi_{*}\approx\phi_{cmb}=\left(\underbrace{\frac{\phi_{e}}{M_p}}_{\leq 1}+\underbrace{\frac{\Delta\phi}{M_p}}_{\leq 1}\right)M_p\leq M_p.\ee
Further applying convergence criteria one can recast Eq~(\ref{mat4}) as:
\begin{equation}\label{mat5}
\left(\begin{tabular}{c}
      $V_{0}$\\
      %$0$\\
      $m^{2}$\\
      %$0$\\
      $24\lambda$\\
      $\cdots$\\
      $\cdots$\\
$\cdots$\\
$\cdots$\\
\end{tabular}\right)
\begin{tabular}{c}
      \\
      $\approx$\\
      \\
      \end{tabular}
\left(\begin{tabular}{c}
      $V(\phi_{*})$\\
      %$V^{'}(\phi_{*})$\\
      $V^{''}(\phi_{*})$\\
     %$V^{'''}(\phi_{*})$\\
      $V^{''''}(\phi_{*})$\\
      $\cdots$\\
      $\cdots$\\
$\cdots$\\
$\cdots$\\
      \end{tabular}\right).
\end{equation}
The present analysis clearly shows that the scale of inflation is given by:
\be\label{scale}
\sqrt[4]{{\cal V}_{inf}}\approx\sqrt[12]{2\pi^2 P_{S}(k_{*})r(k_{*})}M^{1/3}_{p}\sigma^{1/6}\lesssim \sqrt[4]{\frac{3}{2}P_{S}(k_{\star})r(k_{\star})\pi^{2}}~M_p
\ee
Now using Eq~(\ref{mat3}) and Eq~(\ref{mat5}) along with Eq~(\ref{scale}) here I get the following expression for the analytical bound
 on the positive brane tension $\sigma$ in terms of inflationary observables in RS single braneworld setup as:
\be\begin{array}{llll}\label{efolda1aa1}
\underline{\bf Without~~Z_2}:~~~~~
\displaystyle \sigma\leq 
\displaystyle
\left|\frac{3456P_{S}(k_*)\pi^{2}M^{4}_{p}}{(\Delta N_{b})^6 (r(k_*))^2}\right|,\end{array}\ee
\be\begin{array}{llll}\label{efolda2aa1}
\underline{\bf With~~Z_2}:~~~~~
\displaystyle\sigma\leq\displaystyle \left|\frac{3\sqrt{3}P_{S}(k_*)r(k_*)\pi^2 M^4_p}{4\Delta N_b \left(n_{S}(k_*)-1+\frac{r(k_*)}{4}\right)}\right|.\end{array}\ee

where $\phi_{e}\leq M_p$ have been used in Eq~(\ref{efolda2aa1}). 

Further using Eq~(\ref{efolda1aa1}), Eq~(\ref{efolda2aa1}) and Eq~(\ref{mass}) it is possible 
to write down the analytical expression for the upper bound of the 5D Planck mass in terms of 4D Planck mass and various inflationary observables as:
\be\begin{array}{llll}\label{efolda1aa1z}
\underline{\bf Without~~Z_2}:~~~~~
\displaystyle M_5\leq 
\displaystyle
\sqrt[6]{\left|\frac{2\sqrt{2r(k_*)}P_{S}(k_*)\pi^{3}}{\Delta N_{b}}\right|}~M_p,\end{array}\ee
\be\begin{array}{llll}\label{efolda2aa1z}
\underline{\bf With~~Z_2}:~~~~~
\displaystyle M_5\leq\displaystyle \sqrt[6]{\left|\frac{P_{S}(k_*)r(k_*)\pi^3}{\Delta N_b \left(n_{S}(k_*)-1+\frac{3r(k_*)}{8}\right)}\right|}~M_p.\end{array}\ee

Finally using Eq~(\ref{lam}), Eq~(\ref{lamc}) and Eq~(\ref{efolda1aa1}-\ref{efolda2aa1z}) it is possible 
to write down the analytical expression for the upper bound on the magnitude of 5D bulk cosmological constant in terms of 4D Planck mass and various inflationary observables as:
\be\begin{array}{llll}\label{efq}
\underline{\bf Without~~Z_2}:~~~~~
\displaystyle \tilde{\Lambda}_{5}= \frac{\Lambda_{5}}{32\pi}\geq 
\displaystyle -\frac{9}{48}
\sqrt{\left|\frac{(2r(k_*))^{3/2}P^3_{S}(k_*)\pi^{5}}{2(\Delta N_{b})^3}\right|}~M^5_p,\end{array}\ee
\be\begin{array}{llll}\label{efq1}
\underline{\bf With~~Z_2}:~~~~~
\displaystyle \tilde{\Lambda}_{5}= \frac{\Lambda_{5}}{32\pi}\geq\displaystyle -\frac{9}{384}\sqrt{\left|\frac{P^3_{S}(k_*)r^3(k_*)\pi^5}{(\Delta N_b)^3 
\left(n_{S}(k_*)-1+\frac{3r(k_*)}{8}\right)^3}\right|}~M^5_p.\end{array}\ee

Within Planck's observable region of  $\Delta {N}_{b}\sim {\cal O}(8-10)$,  
 it is possible to constrain the power spectrum: $P_S$, spectral tilt: $n_S$,
running of the spectral tilt: $\alpha_S$, and running of running of the spectral tilt: $\kappa_S$, 
for {\bf Planck+WMAP-9+high~L+BICEP2} data sets \cite{Planck-1,Planck-infl}:
 \begin{eqnarray}\label{obscons1}
 0.15 &\leq & r_{b}(k_*)\leq  0.27\\
 \ln(10^{10}P_{S})&=&3.089^{+0.024}_{-0.027}~~ ({\rm within}~ 2\sigma ~C.L.),\\
 n_{S}&=&0.9600 \pm 0.0071~~ ({\rm within}~ 3\sigma ~C.L.),\\
 \alpha_{S}&=&dn_{S}/d\ln k=-0.022\pm 0.010~~({\rm within}~1.5\sigma~C.L.),\\
 \kappa_{S}&=&d^{2}n_{S}/d\ln k^{2}=0.020^{+0.016}_{-0.015}~~({\rm within}~1.5\sigma~C.L.)\,.
 \end{eqnarray}
and for {\bf Planck+WMAP-9+high~L} data sets \cite{Ade:2014xna}:
 \begin{eqnarray}\label{obscons2}
 && r_{b}(k_*)<  0.12\\
 \ln(10^{10}P_{S})&=&3.089^{+0.024}_{-0.027}~~ ({\rm within}~ 2\sigma ~C.L.),\\
 n_{S}&=&0.9603 \pm 0.0073~~ ({\rm within}~ 3\sigma ~C.L.),\\
 \alpha_{S}&=&dn_{S}/d\ln k=-0.013\pm 0.009~~({\rm within}~1.5\sigma~C.L.),\\
 \kappa_{S}&=&d^{2}n_{S}/d\ln k^{2}=0.020^{+0.016}_{-0.015}~~({\rm within}~1.5\sigma~C.L.)\,.
 \end{eqnarray}
which will fix the field excursion in a sub-Planckian region by putting required constraint on the positive brane tension $\sigma$ as discussed earlier. 
Now using these combined constraints it is possible to estimate the approximated numerical bound of the various parameters- brane tension ($\sigma$),
 5D Planck mass ($M_5$) and 5D cosmological constant ($\tilde{\Lambda}_{5}$) lying
 within the following window~\footnote{In order to
recover the observational successes of general relativity, the high-energy regime where significant
deviations occur must take place before nucleosynthesis.
Table-top tests of Newton's laws put the lower bound on the brane tension and 5D Planck
 scale as: $\sigma> {\cal O}(2.86\times 10^{-86})~M^4_p$ and $M_5>{\cal O}(4.11\times 10^{-11})~M_p$. But such lower bound will not be able to
 produce large tensor-to-scalar ratio as required by BICEP2 and the upper bound of Planck.}: 
\be\begin{array}{llll}\label{efqzx}
\large \underline{\bf Without~~Z_2}:~~~~~%\\
%\underline{\bf Using~~Planck+WMAP-9+high~L+BICEP2}:\\
\displaystyle \sigma\leq {\cal O}(10^{-9})~M^4_p,~~~M_5\leq {\cal O}(0.04)~M_p,~~~\tilde{\Lambda}_{5}\geq -{\cal O}(
10^{-15})~M^5_p,%\\
%\underline{\bf Using~~Planck+WMAP-9+high~L}:\\\displaystyle \sigma< {\cal O}(2.08\times 10^{-9})~M^4_p,~~~M_5< {\cal O}(0.04)~M_p,
%~~~\tilde{\Lambda}_{5}\geq -{\cal O}(3.36-3.86)
%\times 10^{-15}~M^5_p, 
\end{array}\ee
\be\begin{array}{llll}\label{efq1zx}
\large\underline{\bf With~~Z_2}:~~~~~%\\
%\underline{\bf Using~~Planck+WMAP-9+high~L+BICEP2}:\\
\displaystyle \sigma\leq {\cal O}(10^{-9})~M^4_p,~~~M_5\leq {\cal O}(0.05)~M_p,~~~\tilde{\Lambda}_{5}\geq -{\cal O}(
10^{-15})~M^5_p.%\\
%\underline{\bf Using~~Planck+WMAP-9+high~L}:\\ \displaystyle \sigma< {\cal O}(4.64\times 10^{-9})~M^4_p,~~~M_5< {\cal O}(0.07)~M_p,
%~~~\tilde{\Lambda}_{5}\geq -{\cal O}(2.03-2.34)
%\times 10^{-15}~M^5_p.
\end{array}\ee
Also I get the following bound on the suppression pre-factor as appearing in the right side of Eq~(\ref{rk9}):
\be\begin{array}{llll}\label{boundf}
   \Large \displaystyle \frac{1}{2}\sqrt{\frac{\sigma}{3{\cal V}_{inf}}}< {\cal O}(0.09-0.16).
   \end{array}\ee

Substituting all of these contributions stated in Eq~(\ref{wq1}-\ref{wq3}) to Eq~(\ref{rk9}) and further using Eq~(\ref{obscons1},\ref{obscons2},\ref{boundf})
 the upper bound of the field excursion ($|\Delta\phi|$) is constrained within the following sub-Planckian regime~\footnote{In the case of
 single field models in four dimensions, assuming the monotonous behaviour of the slow-roll parameter during inflation it has been shown that the tensor-to-scalar ratio
$r>0.1$ requires field excursions that are very very close to or above the Planck scale cut-off \cite{Bramante:2014rva,Antusch:2014cpa}, which is completely in agreement with the well 
known {\it Lyth bound} \cite{Lyth:1996im}. But if the slow-roll parameters follow non-monotonous behaviour during inflation \cite{Choudhury:2013jya,Choudhury:2014sxa,Choudhury:2014uxa,Hotchkiss:2011gz}, then by modifying the power law 
parameterization of the primordial power spectrum 
in presence of running and running of the running of the running of spectral tilt of the power spectrum it is possible to generate tensor-to-scalar 
ratio $r>0.1$ from sub-Planckian field excursion within the framework of Effective Field Theory \cite{Choudhury:2014kma,Choudhury:2014wsa}. In this work, I have explicitly shown that 
in context of 
Randall Sundrum single brane cosmological setup,  by tuning the brane tension in the high density/ high energy regime it is possible to generate tensor-to-scalar ratio $r>0.1$
, provided the constrained value of field excursion is lesser compared to the result available in case of low density/ low energy regime in the braneworld and the 
model independent analysis validates the Effective Field Theory prescription more compared the case discussed in \cite{Choudhury:2014kma,Choudhury:2014wsa}. 
In the low density/ low energy regime Randall Sundrum single braneworld model exactly goes to the General Relativistic limit and 
hence it is possible to achieve the stringent bound derived in 
\cite{Choudhury:2014kma,Choudhury:2014wsa}. On the other hand in the high density regime of the Randall Sundrum braneworld without modifying the power-law/scale invariant parametrization
of the primordial power spectrum it is possible to achieve $r>0.1$ with sub-Planckian field excursion just by allowing fine tuning in the brane tension. But
if we still modify the primordial power spectrum and allow the contributions from running and running of the running of the spectral tilt
 of the primordial power spectrum 
then we can probe lesser value of the field excursion compared to the General Relativistic limiting (low density) result
 derived in \cite{Choudhury:2014kma,Choudhury:2014wsa}. Additionally in the high density regime of Randall Sundrum braneworld by allowing fine-tuning
in the brane tension, it is possible to increase the upper bound of the energy scale of inflation and at best it is possible to achieve the upper bound of tensor-to-scalar ratio as observed by
BICEP2 i.e. $r\sim 0.27$ within the Effective Field Theoretic regime of inflation. See Eq~(\ref{aq1}) for details. However, the recent joint analysis performed by Planck mission along with BICEP2/Keck Array team \cite{Ade:2015tva} and Planck 2015 data \cite{Ade:2015lrj} fix
the upper bound of tensor-to-scalar ratio at $r<0.12$, which can be surely achieved by the prescribed methodology established within the framework of Randall Sundrum single brane inflationary scenario.
}:
\begin{widetext}
\be\begin{array}{lll}\label{rk9fin}
 \displaystyle\left|\frac{\Delta\phi}{M_p}\right|\displaystyle \leq \overbrace{\underbrace{\frac{1}{2}\sqrt{\frac{\sigma}{3{\cal
 V}_{inf}}}}_{{\cal O}(0.09-0.16)}}^{\rm \bf Tuning~factor~
in~RS}\times\underbrace{\left\{\begin{array}{ll}
                    \displaystyle  {\cal O}(2.7-5.1) %&
 %\mbox{\small {\bf for \underline{Case I}}}  
\\ 
         \displaystyle  {\cal O}(2.7-4.6) %& \mbox{\small {\bf for \underline{Case II}}}
\\ 
\displaystyle  {\cal O}(0.6-1.8) %& \mbox{\small {\bf for \underline{Case III}}}
\\ 
\displaystyle  {\cal O}(0.2-0.3) %& \mbox{\small {\bf for \underline{Case IV}}}.
          \end{array}
\right.}_{\rm \bf From~ low~ density~ regime~of~RS}
= \underbrace{\left\{\begin{array}{ll}
                    \displaystyle  {\cal O}(0.24-0.81) &
 \mbox{\small {\bf for \underline{Case I}}}  \\ 
         \displaystyle  {\cal O}(0.24-0.73) & \mbox{\small {\bf for \underline{Case II}}}\\ 
\displaystyle  {\cal O}(0.05-0.28) & \mbox{\small {\bf for \underline{Case III}}}\\ 
\displaystyle  {\cal O}(0.02-0.05) & \mbox{\small {\bf for \underline{Case IV}}}
          \end{array}
\right.}_{\rm \bf From~ high~ density~regime~of~RS}
\end{array}\ee
\end{widetext}
which is consistent with all the observational constraints mentioned earlier.
Now in the low energy regime when the energy density of inflaton $\rho<<\sigma$ then,
in this limit, the suppression pre-factor turns out to be: 
\be \lim_{\rho<<\sigma}\left[\frac{1}{2}\sqrt{\frac{\sigma}{3{\cal V}_{inf}}}\right]\rightarrow 1.\ee 
Using this limiting result it is possible to obtain also
the relation between field excursion and tensor-to-scalar ratio from Eq~(\ref{rk9}) in case of usual GR prescribed effective field theory setup.
For the details see the refs.~\cite{Choudhury:2013iaa,Choudhury:2014kma,Choudhury:2014wsa,Choudhury:2013woa,Choudhury:2014hua} 
where such limit and their cosmological consequences are elaborately studied. Now let me concentrate on the 
{\bf first case} of Eq~(\ref{rk9}), which is the most simplest physical situation.
If I take the limit, $\rho<<\sigma$, then it absolutely reduces to the good-old {\it Lyth bound} in which for $\Delta N_{b}\sim {\cal O}(8-10)$ super-Planckian field excursion 
$|\Delta\phi|\sim {\cal O}(2.7-5.1)~M_p>M_p$ is required
to generate large tensor-to-scalar ratio as observed by BICEP2 or at least generates the tensor-to-scalar ratio consistent with the upper bound of Planck. Now in the 
RS single braneworld setup by setting the brane tension in the above mentioned desired value and fixing the scale of inflation in the vicinity of GUT scale
it is possible to generate large tensor-to-scalar ratio using sub-Planckian field excursion for which it is possible to describe the setup by using effective field theory
of inflation. But only in the {\bf last case} of Eq~(\ref{rk9}) in the limit $\rho<<\sigma$ it is possible to obtain 
sub-Planckian field excursion $|\Delta\phi|\sim {\cal O}(0.2-0.3)~M_p<M_p$ to get large value of tensor-to-scalar ratio \cite{Choudhury:2013iaa,Choudhury:2014kma,Choudhury:2014wsa}. If we now switch on the effect of single brane in RS
setup then due to the presence of the suppression pre-factor as mentioned in Eq~(\ref{boundf}) the field excursion further reduces to the GUT scale i.e. 
$|\Delta\phi|\sim {\cal O}(0.02-0.05)~M_p<M_p$.

\section{Conclusion}
To summarize, in the present article, I have established a methodology for generating sub-Planckian field excursion along with large tensor-to-scalar ratio 
in a single brane RS braneworld scenario for generic model of inflation with and without ${\bf Z_{2}}$ symmetry in the most generalized form of 
inflationary potential. I have investigated this scenario by incorporating various parametrization in the power spectrum for scalar and tensor modes
as well as in the tensor-to-scalar ratio as required by the observational probes. Using the proposed technique I have further derived a analytical as well as the 
numerical constraints on the positive brane tension, 5D Planck scale and 5D bulk cosmological constant in terms of the 4D Planck scale. Finally, I have given
 an estimation of the field excursion which lies within a sub-Planckian regime and makes the embedding of inflationary paradigm in RS single braneworld via 
effective field theory prescription consistent.  

\section*{Acknowledgments:} 
 I would like to thank Council of Scientific and
Industrial Research, India for financial support through Senior
Research Fellowship (Grant No. 09/093(0132)/2010) and Department of Theoretical Physics, Tata Institute of Fundamental Research, Mumbai for providing me Visiting (Post-Doctoral) Research Fellowship. 
 I take this opportunity to thank sincerely to Prof. Soumitra SenGupta, Dr. Anupam Mazumdar and Dr. Supratik Pal for their constant support and inspiration.
%%%%%%%%%%%%%%%%%%%%%%%%%%%%%%%%%%%%%%%%%%%%%%%%%%%%%%%%%%%%%%%%%%%%%%%%%%%%%%%%%%%%%%%%%%%%%%%%%%%%%%%%%%%%%%%%%%%%%%%%%%%%%%%%%%%%%%%%%%%%%%%%%%%%%%%%%%%%%%%%%%%%%%%%%%%%%%%%%%%%%%%%%%%%%%%%%%%%%%%%%%%%%%%%%%%%%%%%%%%%%
%%%%%%%%%%%%%%%%%%%%%%%%%%%%%%%%%%%%%%%%%%%%%%%%%%%%%%%%%%%%%%%%%%%%%%%%%%%%%%%%%%%%%%%%%%%%%%%%%%%%%%%%%%%%%%%%%%%%%%%%%%%%%%%%%%%%%%%%%%%%%%%%%%%%%%%%%%%%%%%%%%%%%%%%%%%%%%%%%%%%%%%%%%%%%%%%%%%%%%%%

%\section*{Acknowledgments}

%%%%%%%%%%%%%%%%%%%%%%%%%%%%%%%%%%%%%%%%%%%%%%%%%%%%%%%%%%%%%%%%%%%%%%%%%%%%%%%%%%%%%%%%%%%%%%%%%%%%%%%%%%%%%%%%%%%%%%%%%%%%%%%%%%%%%%%%%%%%%%%%%%%%%%%%%%%%%%%%%%%%%%%%%%%%%%%%%%%%%%%%
%%%%%%%%%%%%%%%%%%%%%%%%%%%%%%%%%%%%%%%%%%%%%%%%%%%%%%%%%%%%%%%%%%%%%%%%%%%%%%%%%%%%%%%%%%%%%%%%%%%%%%%%%%%%%%%%%%%%%%%%%%%%%%%%%%%%%%%%%%%%%%%%%%%%%%%%%%%%%%%%%%%%%%%%%%%%%%%%%%%%%%%%%%%%%%%%%%%
\section*{Appendix} 

\subsection*{\bf A. Consistency relations in RS single braneworld}

In the context of RS single braneworld the spectral indices $(n_S, n_T)$, running $(\alpha_S, \alpha_T)$ and running of the running $(\kappa_T,\kappa_S)$ at the momentum pivot scale $k_*$ 
can be expressed as \cite{Maartens:2010ar,Choudhury:2011sq,Choudhury:2012ib}:
\begin{eqnarray}
 n_S(k_*) -1&=& 2\eta_{b}(\phi_*)-6\epsilon_{b}(k_*),\\
n_T(k_*) &=& -3\epsilon_{b}(k_*)=-\frac{r_{b}(k_*)}{8},\\
\alpha_S(k_*) &=& 16\eta_{b}(k_*)\epsilon_{b}(k_*)-18\epsilon^{2}_{b}(k_*)-2\xi^{2}_{b}(k_*),\\
\alpha_T(k_*) &=& 6\eta_{b}(k_*)\epsilon_{b}(k_*)-9\epsilon^{2}_{b}(k_*),\\
\kappa_S(k_*)&=&152\eta_{b}(k_*)\epsilon^{2}_{b}(k_*)-32\epsilon_{b}(k_*)\eta^{2}_{b}(k_*)-108\epsilon^{3}_{b}(k_*)
-24\xi^{2}_{b}(k_*)\epsilon_{b}(k_*)+2\eta_{b}(k_*)\xi^{2}_{b}(k_*)+2\sigma^{3}_{b}(k_*),\\
\kappa_T(k_*)&=&66\eta_{b}(k_*)\epsilon^{2}_{b}(k_*)-12\epsilon_{b}(k_*)\eta^{2}_{b}(k_*)
-54\epsilon^{3}_{b}(k_*)-6\epsilon_{b}(k_*)\xi^{2}_{b}(k_*).
\end{eqnarray}
 Let me compute the following significant contributions which appeared at the left side of Eq~(\ref{rk9}) in terms of slow-roll parameters in RS single braneworld:
\begin{eqnarray}
 \label{wq1}n_T(k_*)-n_S(k_*)+1&=&\left(\frac{d\ln r_{b}(k)}{d\ln k}\right)_*=\left[\frac{r_{b}(k_*)}{8}-2\eta_{b}(k_*)\right],\\
\label{wq2}\alpha_T(k_*)-\alpha_S(k_*)&=&\left(\frac{d^2\ln r_{b}(k)}{d\ln k^2}\right)_*=\left[\left(\frac{r_{b}(k_*)}{8}\right)^2-\frac{20}{3}\left(\frac{r_{b}(k_*)}{8}\right)+2\xi^2_{b}(k_*)\right],\\
\label{wq3}\kappa_T(k_*)-\kappa_S(k_*)&=&\left(\frac{d^3\ln r_{b}(k)}{d\ln k^3}\right)_* \nonumber\\&=&
\left[2\left(\frac{r_{b}(k_*)}{8}\right)^3-\frac{86}{9}\left(\frac{r_{b}(k_*)}{8}\right)^2
+\frac{4}{3}\left(6\xi^2_{b}(k_*)+5\eta^{2}_{b}(k_*)\right)\left(\frac{r_{b}(k_*)}{8}\right)
+2\eta_{b}(k_*)\xi^2_{b}(k_*)+2\sigma^{3}_{b}(k_*)\right]\nonumber\\.
\end{eqnarray}
Here Eq~(\ref{wq1}-\ref{wq3})) represent the running, running of the running and running of the double running of tensor-to-scalar ratio.
\newpage
%%%%%%%%%%%%%%%%%%%%%%%%%%%%%%%%%%%%%%%%%%%%%%%%%%%%%%%%%%%%%%%%%%%%%%%%%%%%%%%%%%%%%%%%%%%%%%%%%%%%%%%%%%%%%%%%%%%%%%%%%%%%%%%%%%%%%%%%%%%%%%%%%%%%%%%%%%%%%%%%%%%

\subsection*{\bf B. Computation of Momentum integral}

Now let us explicitly compute left hand side of the Eq~(\ref{con2}). To serve this purpose I start with the computation of momentum integration where I investigate the possibility of four physical 
situations as mentioned in Eq~(\ref{rk1}) finally leading to: 
\begin{widetext}
\be\begin{array}{lll}\label{rk2}
 \displaystyle\int^{k_{cmb}}_{k_{e}}d\ln k ~\sqrt{r_{b}(k)}\displaystyle =\left\{\begin{array}{ll}
                    \displaystyle  \sqrt{r_{b}(k_*)}\ln\left(\frac{k_{cmb}}{k_{e}}\right) &
 \mbox{\small {\bf for \underline{Case I}}}  \\ 
         \displaystyle  \frac{2\sqrt{r_{b}(k_*)}}{n_{T}(k_{*})-n_{S}(k_{*})+1}\left[\left(\frac{k_{cmb}}{k_{*}}\right)^{\frac{n_{T}(k_{*})-n_{S}(k_{*})+1}{2}}-\left(\frac{k_{e}}{k_{*}}\right)^{\frac{n_{T}(k_{*})-n_{S}(k_{*})+1}{2}}\right] & \mbox{\small {\bf for \underline{Case II}}}\\
\displaystyle  \sqrt{r_{b}(k_*)}e^{-\frac{(n_{T}(k_{*})-n_{S}(k_{*})+1)^2}{2(\alpha_{T}(k_{*})-\alpha_{S}(k_{*}))}}\sqrt{\frac{2\pi}{(\alpha_{T}(k_{*})-\alpha_{S}(k_{*}))}}
\\ \displaystyle \left[{\rm erfi}\left(\frac{n_{T}(k_{*})-n_{S}(k_{*})+1}{\sqrt{
2(\alpha_{T}(k_{*})-\alpha_{S}(k_{*}))}}+\sqrt{\frac{(\alpha_{T}(k_{*})-\alpha_{S}(k_{*}))}{8}}\ln\left(\frac{k_{cmb}}{k_{*}}\right)\right)
\right.\\ \left. \displaystyle ~~~~~~
-{\rm erfi}\left(\frac{n_{T}(k_{*})-n_{S}(k_{*})+1}{\sqrt{2(\alpha_{T}(k_{*})-\alpha_{S}(k_{*}))}}+\sqrt{\frac{(\alpha_{T}(k_{*})-\alpha_{S}(k_{*}))}{8}}\ln\left(\frac{k_{e}}{k_{*}}\right)\right)\right] & \mbox{\small {\bf for \underline{Case III}}}\\ 
\displaystyle  \sqrt{r_{b}(k_{*})}\left[\left(\frac{3}{2}-\frac{n_{T}(k_{*})-n_{S}(k_{*})}{2}+\frac{\alpha_{T}(k_{*})-\alpha_{S}(k_{*})}{8}\right.\right.\\ \left.\left. 
\displaystyle ~~~~~\displaystyle -\frac{\kappa_{T}(k_{*})-\kappa_{S}(k_{*})}{24}\right)\displaystyle \left\{\frac{k_{cmb}}{k_{*}}-\frac{k_{e}}{k_{*}}\right\}-
\left(\frac{1}{2}-\frac{n_{T}(k_{*})-n_{S}(k_{*})}{2}\right.\right.\\ \left.\left.\displaystyle \displaystyle+\frac{\alpha_{T}(k_{*})-\alpha_{S}(k_{*})}{8} 
 -\frac{\kappa_{T}(k_{*})-\kappa_{S}(k_{*})}{24}\right)\displaystyle \left\{\frac{k_{cmb}}{k_{*}}\ln\left(\frac{k_{cmb}}{k_{*}}\right)
-\frac{k_{e}}{k_{*}}\ln\left(\frac{k_{e}}{k_{*}}\right)\right\}\right.\\ \left.\displaystyle +\left(\frac{\kappa_{T}(k_{*})
-\kappa_{S}(k_{*})}{48}-\frac{\alpha_{T}(k_{*})-\alpha_{S}(k_{*})}{16}\right)\displaystyle\left\{\frac{k_{cmb}}{k_{*}}\ln^2\left(\frac{k_{cmb}}{k_{*}}\right)
-\frac{k_{e}}{k_{*}}\ln^2\left(\frac{k_{e}}{k_{*}}\right)\right\}\right.\\ \left.\displaystyle 
-\frac{\kappa_{T}(k_{*})
-\kappa_{S}(k_{*})}{144}\displaystyle\left\{\frac{k_{cmb}}{k_{*}}\ln^3\left(\frac{k_{cmb}}{k_{*}}\right)
-\frac{k_{e}}{k_{*}}\ln^3\left(\frac{k_{e}}{k_{*}}\right)\right\}\right] & \mbox{\small {\bf for \underline{Case IV}}}.
          \end{array}
\right.
\end{array}\ee
\end{widetext}

where in a realistic physical situation one assumes the pivot scale of momentum $k_{*}\approx k_{cmb}$.
Now further substituting Eq~(\ref{intnk1}) on Eq~(\ref{rk2}) I get:
  \begin{widetext}
\be\begin{array}{lll}\label{rk3}
 \displaystyle\int^{k_{cmb}}_{k_{e}}d\ln k ~\sqrt{r_{b}(k)}\displaystyle =\left\{\begin{array}{ll}
                    \displaystyle  \sqrt{r_{b}(k_*)}\Delta {N}_{b} &
 \mbox{\small {\bf for \underline{Case I}}}  \\ 
         \displaystyle  \frac{2\sqrt{r_{b}(k_*)}}{n_{T}(k_{*})-n_{S}(k_{*})+1}\left[1-e^{-\Delta {N}_{b}\left(\frac{n_{T}(k_{*})-n_{S}(k_{*})+1}{2}\right)}\right] & \mbox{\small {\bf for \underline{Case II}}} \\
\displaystyle  \sqrt{r_{b}(k_*)}e^{-\frac{(n_{T}(k_{*})-n_{S}(k_{*})+1)^2}{2(\alpha_{T}(k_{*})-\alpha_{S}(k_{*}))}}\sqrt{\frac{2\pi}{(\alpha_{T}(k_{*})-\alpha_{S}(k_{*}))}}
\\ \displaystyle \left[{\rm erfi}\left(\frac{n_{T}(k_{*})-n_{S}(k_{*})+1}{\sqrt{
2(\alpha_{T}(k_{*})-\alpha_{S}(k_{*}))}}\right)
\right.\\ \left. \displaystyle ~~~~~~
-{\rm erfi}\left(\frac{n_{T}(k_{*})-n_{S}(k_{*})+1}{\sqrt{2(\alpha_{T}(k_{*})-\alpha_{S}(k_{*}))}}-\sqrt{\frac{(\alpha_{T}(k_{*})-\alpha_{S}(k_{*}))}{8}}
\Delta {N}_{b}\right)\right] & \mbox{\small {\bf for \underline{Case III}}}\\ 
\displaystyle  \sqrt{r_{b}(k_{*})}\left[\left(\frac{3}{2}-\frac{n_{T}(k_{*})-n_{S}(k_{*})}{2}+\frac{\alpha_{T}(k_{*})-\alpha_{S}(k_{*})}{8}\right.\right.\\ \left.\left. 
\displaystyle ~~~~~\displaystyle -\frac{\kappa_{T}(k_{*})-\kappa_{S}(k_{*})}{24}\right)\displaystyle \left\{1-e^{-\Delta {N}_{b}}\right\}-
\left(\frac{1}{2}-\frac{n_{T}(k_{*})-n_{S}(k_{*})}{2}\right.\right.\\ \left.\left.\displaystyle \displaystyle+\frac{\alpha_{T}(k_{*})-\alpha_{S}(k_{*})}{8} 
 -\frac{\kappa_{T}(k_{*})-\kappa_{S}(k_{*})}{24}\right)\displaystyle 
\Delta {N}_{b}e^{-\Delta {N}_{b}}\right.\\ \left.\displaystyle -\left(\frac{\kappa_{T}(k_{*})
-\kappa_{S}(k_{*})}{48}-\frac{\alpha_{T}(k_{*})-\alpha_{S}(k_{*})}{16}\right)\displaystyle
(\Delta {N}_{b})^{2}e^{-\Delta {N}_{b}}\right.\\ \left.\displaystyle 
-\frac{\kappa_{T}(k_{*})
-\kappa_{S}(k_{*})}{144}\displaystyle
(\Delta {N}_{b})^{3}e^{-\Delta {N}_{b}}\right] & \mbox{\small {\bf for \underline{Case IV}}}.
          \end{array}
\right.
\end{array}\ee
\end{widetext}
Now for completeness let me concentrate on a limiting situation where $\Delta {N}_{b}$ is small but within the observable range. In such a situation one has the following results: 
  \begin{widetext}
\be\begin{array}{lll}\label{rk4}
 \displaystyle\lim_{\Delta {N}_{b}\rightarrow small}\left[\int^{k_{cmb}}_{k_{e}}d\ln k ~\sqrt{r_{b}(k)}\right]\displaystyle =\left\{\begin{array}{ll}
                    %\displaystyle  \sqrt{r_{b}(k_*)}\Delta {N}_{b} &
 %\mbox{\small {\bf for \underline{Case I}}}  \\ \\
         \displaystyle  \sqrt{r_{b}(k_*)}\Delta {N}_{b} & \mbox{\small {\bf for \underline{Case II}}}\\ 
\displaystyle  \sqrt{r_{b}(k_*)}\Delta {N}_{b}e^{-\frac{(n_{T}(k_{*})-n_{S}(k_{*})+1)^2}{2(\alpha_{T}(k_{*})-\alpha_{S}(k_{*}))}} & \mbox{\small {\bf for \underline{Case III}}} \\
\displaystyle  \sqrt{r_{b}(k_{*})}\Delta {N}_{b}\left[1-\left(\frac{\kappa_{T}(k_{*})
-\kappa_{S}(k_{*})}{48}-\frac{\alpha_{T}(k_{*})-\alpha_{S}(k_{*})}{16}\right)\displaystyle
\Delta {N}_{b}\right.\\ \left.\displaystyle 
~~~~~~~~~~~~~~~~~~~~~~~~~~~~~~~~~~~~~~~~~~-\frac{\kappa_{T}(k_{*})
-\kappa_{S}(k_{*})}{144}\displaystyle
(\Delta {N}_{b})^{2}\right] & \mbox{\small {\bf for \underline{Case IV}}}.
          \end{array}
\right.
\end{array}\ee
\end{widetext}
%%%%%%%%%%%%%%%%%%%%%%%%%%%%%%%%%%%%%%%%%%%%%%%%%%%%%%%%%%%%%%%%%%%%%%%%%%%%%%%%%%%%%%%%%%%%%%%%%%%%%%%%%%%%%%%%%%%%%%%%%%%%%%%%%%%%%%%%%%%%%%%%%%%%%%%%%%%%%%%%%%%

\subsection*{\bf C. Computation of Potential dependent integral}

Next I compute the right hand side of the Eq~(\ref{con2}). To serve this purpose 
I start with Eq~(\ref{rt10a}). 
  \begin{widetext}
\be\begin{array}{lll}\label{rk6}
 \displaystyle\int^{\phi_{cmb}}_{\phi_{e}}d\phi ~\sqrt{V(\phi)}\displaystyle =
         \displaystyle  \sqrt{V(\phi_{0})}\left(\frac{\Delta\phi}{M_p}\right)\left[1+\frac{1}{2}\sum^{\infty}_{n=1}\frac{V^{'n}(\phi_0)M^{n}_{p}}{n!V(\phi_0)}
\left(\frac{\phi_{e}-\phi_{0}}{M_{p}}\right)^{n}\right]%,\\
%~~~~~~~~~~~~~~~~~~~~~~~\displaystyle
\approx\sqrt{V(\phi_{0})}\left(\frac{\Delta\phi}{M_p}\right) 
\end{array}\ee
\end{widetext}
where in the next to last step I have used the convergent criteria of the series sum as mentioned earlier in this paper.
Similarly from Eq~(\ref{rt10ab}) I get:
  \begin{widetext}
\be\begin{array}{lll}\label{rk7}
 \displaystyle\int^{\phi_{cmb}}_{\phi_{e}}d\phi ~\sqrt{V(\phi)}\displaystyle =
         \displaystyle  \sqrt{V_{0}}\left(\frac{\Delta\phi}{M_p}\right)\left[1+\frac{1}{2}\sum^{\infty}_{m=1}\frac{{\bf C}_{2m}M^{2m}_{p}}{(2m+1)V_0}
\left(\frac{\phi_{e}-\phi_{0}}{M_{p}}\right)^{2m}\right]%,\\
%~~~~~~~~~~~~~~~~~~~~~~~\displaystyle
\approx\sqrt{V_{0}}\left(\frac{\Delta\phi}{M_p}\right) 
\end{array}\ee
\end{widetext}
Now further clubbing Eq~(\ref{rk6}) and Eq~(\ref{rk7}) with/without ${\bf Z}_{2}$ symmetric physical situation I get:
\begin{widetext}
\be\begin{array}{lll}\label{rk8}
 \displaystyle\int^{\phi_{cmb}}_{\phi_{e}}d\phi ~\sqrt{V(\phi)}\displaystyle 
\approx\sqrt{{\cal V}_{inf}}\left(\frac{\Delta\phi}{M_p}\right).
\end{array}\ee
\end{widetext}
where the scale of inflation is determined by the symbol, ${\cal V}_{inf}=V_{0}$ for $\phi_{0}=0$ and ${\cal V}_{inf}=V(\phi_0)$ for $\phi_0\neq 0$.
\newpage

%%%%%%%%%%%%%%%%%%%%%%%%%%%%%%%%%%%%%%%%%%%%%%%%%%%%%%%%%%%%%%%%%%%%%%%%%%%%%%%%%%%%%%%%%%%%%%%%%%%%%%%%%%%%%%%%%%%%%%%%%%%%%%%%%%%%%%%%%%%%%%%%%%%%%%%%%%%%%%%%%%%

\subsection*{\bf D. Relationship between brane tension and the scale of inflation in RS setup}

Further I get the following relationship between brane tension and the scale of inflation:
\begin{widetext}
\be\begin{array}{lll}\label{rk9xc} 
    \underline{\bf Without~~Z_2}:~~~~~\\
 \displaystyle\frac{\sigma}{{\cal V}_{inf}}\displaystyle =\frac{1}{3}\times\left\{\begin{array}{ll}
                    \displaystyle  1 &
 \mbox{\small {\bf for \underline{Case I}}}  \\ 
         \displaystyle  \frac{4}{(n_{T}(k_{*})-n_{S}(k_{*})+1)^2|\Delta N_b|^2}\left|1-e^{-\Delta {N}_{b}\left(\frac{n_{T}(k_{*})-n_{S}(k_{*})+1}{2}\right)}\right|^2 & \mbox{\small {\bf for \underline{Case II}}} \\
\displaystyle  \frac{1}{|\Delta N_b|^2}e^{-\frac{(n_{T}(k_{*})-n_{S}(k_{*})+1)^2}{(\alpha_{T}(k_{*})-\alpha_{S}(k_{*}))}}\left(\frac{2\pi}{(\alpha_{T}(k_{*})-\alpha_{S}(k_{*}))}\right)
\\ \displaystyle \left|{\rm erfi}\left(\frac{n_{T}(k_{*})-n_{S}(k_{*})+1}{\sqrt{
2(\alpha_{T}(k_{*})-\alpha_{S}(k_{*}))}}\right)
\right.\\ \left. \displaystyle ~~~~~~
-{\rm erfi}\left(\frac{n_{T}(k_{*})-n_{S}(k_{*})+1}{\sqrt{2(\alpha_{T}(k_{*})-\alpha_{S}(k_{*}))}}-\sqrt{\frac{(\alpha_{T}(k_{*})-\alpha_{S}(k_{*}))}{8}}
\Delta {N}_{b}\right)\right|^2 & \mbox{\small {\bf for \underline{Case III}}}\\ 
\displaystyle  \frac{1}{|\Delta N_b|^2}\left|\left(\frac{3}{2}-\frac{n_{T}(k_{*})-n_{S}(k_{*})}{2}+\frac{\alpha_{T}(k_{*})-\alpha_{S}(k_{*})}{8}\right.\right.\\ \left.\left. 
\displaystyle ~~~~~\displaystyle -\frac{\kappa_{T}(k_{*})-\kappa_{S}(k_{*})}{24}\right)\displaystyle \left\{1-e^{-\Delta {N}_{b}}\right\}-
\left(\frac{1}{2}-\frac{n_{T}(k_{*})-n_{S}(k_{*})}{2}\right.\right.\\ \left.\left.\displaystyle \displaystyle+\frac{\alpha_{T}(k_{*})-\alpha_{S}(k_{*})}{8} 
 -\frac{\kappa_{T}(k_{*})-\kappa_{S}(k_{*})}{24}\right)\displaystyle 
\Delta {N}_{b}e^{-\Delta {N}_{b}}\right.\\ \left.\displaystyle -\left(\frac{\kappa_{T}(k_{*})
-\kappa_{S}(k_{*})}{48}-\frac{\alpha_{T}(k_{*})-\alpha_{S}(k_{*})}{16}\right)\displaystyle
(\Delta {N}_{b})^{2}e^{-\Delta {N}_{b}}\right.\\ \left.\displaystyle 
-\frac{\kappa_{T}(k_{*})
-\kappa_{S}(k_{*})}{144}\displaystyle
(\Delta {N}_{b})^{3}e^{-\Delta {N}_{b}}\right|^2 & \mbox{\small {\bf for \underline{Case IV}}}.
          \end{array}
\right.
\end{array}\ee
\end{widetext}

\begin{widetext}
\be\begin{array}{lll}\label{rk9xc1} 
    \underline{\bf With~~Z_2}:~~~~~\\
 \displaystyle\frac{\sigma}{{\cal V}_{inf}}\displaystyle =\frac{r(k_*)}{48\left(n_{S}(k_*)-1+\frac{3r(k_*)}{8}\right)^2}\times\left\{\begin{array}{ll}
                    \displaystyle  1 &
 \mbox{\small {\bf for \underline{Case I}}}  \\ 
         \displaystyle  \frac{4}{(n_{T}(k_{*})-n_{S}(k_{*})+1)^2|\Delta N_b|^2}\left|1-e^{-\Delta {N}_{b}\left(\frac{n_{T}(k_{*})-n_{S}(k_{*})+1}{2}\right)}\right|^2 & \mbox{\small {\bf for \underline{Case II}}} \\
\displaystyle  \frac{1}{|\Delta N_b|^2}e^{-\frac{(n_{T}(k_{*})-n_{S}(k_{*})+1)^2}{(\alpha_{T}(k_{*})-\alpha_{S}(k_{*}))}}\left(\frac{2\pi}{(\alpha_{T}(k_{*})-\alpha_{S}(k_{*}))}\right)
\\ \displaystyle \left|{\rm erfi}\left(\frac{n_{T}(k_{*})-n_{S}(k_{*})+1}{\sqrt{
2(\alpha_{T}(k_{*})-\alpha_{S}(k_{*}))}}\right)
\right.\\ \left. \displaystyle ~~~~~~
-{\rm erfi}\left(\frac{n_{T}(k_{*})-n_{S}(k_{*})+1}{\sqrt{2(\alpha_{T}(k_{*})-\alpha_{S}(k_{*}))}}-\sqrt{\frac{(\alpha_{T}(k_{*})-\alpha_{S}(k_{*}))}{8}}
\Delta {N}_{b}\right)\right|^2 & \mbox{\small {\bf for \underline{Case III}}}\\ 
\displaystyle  \frac{1}{|\Delta N_b|^2}\left|\left(\frac{3}{2}-\frac{n_{T}(k_{*})-n_{S}(k_{*})}{2}+\frac{\alpha_{T}(k_{*})-\alpha_{S}(k_{*})}{8}\right.\right.\\ \left.\left. 
\displaystyle ~~~~~\displaystyle -\frac{\kappa_{T}(k_{*})-\kappa_{S}(k_{*})}{24}\right)\displaystyle \left\{1-e^{-\Delta {N}_{b}}\right\}-
\left(\frac{1}{2}-\frac{n_{T}(k_{*})-n_{S}(k_{*})}{2}\right.\right.\\ \left.\left.\displaystyle \displaystyle+\frac{\alpha_{T}(k_{*})-\alpha_{S}(k_{*})}{8} 
 -\frac{\kappa_{T}(k_{*})-\kappa_{S}(k_{*})}{24}\right)\displaystyle 
\Delta {N}_{b}e^{-\Delta {N}_{b}}\right.\\ \left.\displaystyle -\left(\frac{\kappa_{T}(k_{*})
-\kappa_{S}(k_{*})}{48}-\frac{\alpha_{T}(k_{*})-\alpha_{S}(k_{*})}{16}\right)\displaystyle
(\Delta {N}_{b})^{2}e^{-\Delta {N}_{b}}\right.\\ \left.\displaystyle 
-\frac{\kappa_{T}(k_{*})
-\kappa_{S}(k_{*})}{144}\displaystyle
(\Delta {N}_{b})^{3}e^{-\Delta {N}_{b}}\right|^2 & \mbox{\small {\bf for \underline{Case IV}}}.
          \end{array}
\right.
\end{array}\ee
\end{widetext}

In the limiting situation when $\Delta N_{b}$ is small but lies within the observable window,
I get the following relationship between brane tension and the scale of inflation:
\begin{widetext}
\be\begin{array}{lll}\label{rk10zx}
    \underline{\bf Without~~Z_2}:~~~~~\\
 \displaystyle\lim_{\Delta {N}_{b}\rightarrow small}\frac{\sigma}{{\cal V}_{inf}}=\frac{1}{3}\times\left\{\begin{array}{ll}
                   % \displaystyle  \sqrt{r_{b}(k_*)}|\Delta {N}_{b}| &
 %\mbox{\small {\bf for \underline{Case I}}}  \\ \\
         \displaystyle  1 & \mbox{\small {\bf for \underline{Case II}}}\\ 
\displaystyle  e^{-\frac{(n_{T}(k_{*})-n_{S}(k_{*})+1)^2}{(\alpha_{T}(k_{*})-\alpha_{S}(k_{*}))}}\ & \mbox{\small {\bf for \underline{Case III}}}\\ 
\displaystyle  \left|1-\left(\frac{\kappa_{T}(k_{*})
-\kappa_{S}(k_{*})}{48}-\frac{\alpha_{T}(k_{*})-\alpha_{S}(k_{*})}{16}\right)\displaystyle
\Delta {N}_{b}\right.\\ \left.\displaystyle 
~~~~~~~~~~~~~~~~~~~~~~~~~~~~~~~~~~~~~~~~~~-\frac{\kappa_{T}(k_{*})
-\kappa_{S}(k_{*})}{144}\displaystyle
(\Delta {N}_{b})^{2}\right|^2 & \mbox{\small {\bf for \underline{Case IV}}}.
          \end{array}
\right.
\end{array}\ee
\end{widetext}

\begin{widetext}
\be\begin{array}{lll}\label{rk10zx1}
    \underline{\bf With~~Z_2}:~~~~~\\
 \displaystyle\lim_{\Delta {N}_{b}\rightarrow small}\frac{\sigma}{{\cal V}_{inf}}=\frac{r(k_*)}{48\left(n_{S}(k_*)-1+\frac{3r(k_*)}{8}\right)^2}\times\left\{\begin{array}{ll}
                   % \displaystyle  \sqrt{r_{b}(k_*)}|\Delta {N}_{b}| &
 %\mbox{\small {\bf for \underline{Case I}}}  \\ \\
         \displaystyle  1 & \mbox{\small {\bf for \underline{Case II}}}\\ 
\displaystyle  e^{-\frac{(n_{T}(k_{*})-n_{S}(k_{*})+1)^2}{(\alpha_{T}(k_{*})-\alpha_{S}(k_{*}))}}\ & \mbox{\small {\bf for \underline{Case III}}}\\ 
\displaystyle  \left|1-\left(\frac{\kappa_{T}(k_{*})
-\kappa_{S}(k_{*})}{48}-\frac{\alpha_{T}(k_{*})-\alpha_{S}(k_{*})}{16}\right)\displaystyle
\Delta {N}_{b}\right.\\ \left.\displaystyle 
~~~~~~~~~~~~~~~~~~~~~~~~~-\frac{\kappa_{T}(k_{*})
-\kappa_{S}(k_{*})}{144}\displaystyle
(\Delta {N}_{b})^{2}\right|^2 & \mbox{\small {\bf for \underline{Case IV}}}.
          \end{array}
\right.
\end{array}\ee
\end{widetext}
%%%%%%%%%%%%%%%%%%%%%%%%%%%%%%%%%%%%%%%%%%%%%%%%%%%%%%%%%%%%%%%%%%%%%%%%%%%%%%%%%%%%%%%%%%%%%%%%%%%%%%%%%%%%%%%%%%%%%%%%%%%%%%%%%%%%%%%%%%%%%%%%%%%%%%%%%%%%%%%%%%%

\subsection*{\bf E. Computation of analytic expression for $\Delta N_{b}$ in terms of potential}

Let me now compute the analytical expression for $\Delta N_{b}$ using Eq~(\ref{efolda}) and the explicit form of the potential stated in
 Eq~(\ref{rt10a}) and Eq~(\ref{rt10ab}) for consistency check. 
\be\begin{array}{llll}\label{efolda1}
\underline{\bf Without~~Z_2}:~~~~~\\
\displaystyle\Delta N_{b}\approx 
\frac{V^{2}(\phi_0)\Delta\phi}{2\sigma V^{'}(\phi_0)M^2_p}\left[1-\sum^{\infty}_{p=1}\frac{V^{'p}(\phi_0)M^{p-1}_{p}}{(p-1)!V^{'}(\phi_0)}
\left(\frac{\phi_{e}-\phi_0}{M_{p}}\right)^{p-1}
+2\sum^{\infty}_{n=1}\frac{V^{'n}(\phi_0)M^{n}_{p}}{n!V(\phi_0)}\left(\frac{\phi_{e}-\phi_0}{M_{p}}\right)^{n}
%+\sum^{\infty}_{m=1}\frac{V^{'m}(\phi_0)}{m!V(\phi_0)}(\phi-\phi_0)^{m}%\nonumber\right. \\ &&\left.~~
\right. \\ \left.\displaystyle~~~~~~+\sum^{\infty}_{n=1}\sum^{\infty}_{m=1}\frac{V^{'n}(\phi_0)V^{'m}(\phi_0)M^{m+n}_{p}}{n!m!V^{2}(\phi_0)}
\left(\frac{\phi_{e}-\phi_0}{M_{p}}\right)^{n+m}+2\sum^{\infty}_{n=1}\sum^{\infty}_{p=1}\frac{V^{'n}(\phi_0)V^{'p}(\phi_0)M^{n+p-1}_{p}}{n!(p-1)!V(\phi_0)V^{'}(\phi_0)}
\left(\frac{\phi_{e}-\phi_0}{M_{p}}\right)^{n+p-1}
\right. \\ \left.~~~~~~\displaystyle%+\sum^{\infty}_{m=1}\sum^{\infty}_{p=1}\frac{V^{'m}(\phi_0)V^{'p}(\phi_0)M^{m+p-1}_{p}}{n!(p-1)!V(\phi_0)V^{'}(\phi_0)}
%\left(\frac{\phi_{e}-\phi_0}{M_{p}}\right)^{m+p-1}%\nonumber\right. \\ &&\left.~~
+\sum^{\infty}_{n=1}\sum^{\infty}_{m=1}\sum^{\infty}_{p=1}\frac{V^{'n}(\phi_0)V^{'m}(\phi_0)V^{'p}(\phi_0)M^{n+m+p-1}_{p}}{n!m!(p-1)!V^{2}(\phi_0)V^{'}(\phi_0)}
\left(\frac{\phi_{e}-\phi_0}{M_{p}}\right)^{n+m+p-1}\right]\approx \displaystyle
\frac{V^{2}(\phi_0)\Delta\phi}{2\sigma V^{'}(\phi_0)M^2_p},\end{array}\ee
\be\begin{array}{llll}\label{efolda2}
\underline{\bf With~~Z_2}:~~~~~\\
\displaystyle\Delta N_{b}\approx\frac{V^{2}_{0}\Delta\phi}{4\sigma m^{2}M^{2}_{p}}\left[\frac{1}{\Delta\phi}\ln\left(1+\frac{\Delta\phi}{\phi_{e}}\right)
+2\sum^{\infty}_{m=1}\frac{{\bf C}_{2m}M^{2m}_{p}}{V_0}\left(\frac{\phi_{e}}{M_p}\right)^{2m}-
\sum^{\infty}_{p=2}\frac{p{\bf C}_{2p}M^{2p-2}_{p}}{m^{2}}\left(\frac{\phi_{e}}{M_p}\right)^{2p-2}\right.\\ \left.~~~~~~~~~~\displaystyle 
+\sum^{\infty}_{m=1}\sum^{\infty}_{n=1}\frac{{\bf C}_{2m}{\bf C}_{2n}M^{2(n+m)}_{p}}{V^{2}_0}\left(\frac{\phi_{e}}{M_p}\right)^{2(n+m)}
-2\sum^{\infty}_{m=1}\sum^{\infty}_{p=2}\frac{p{\bf C}_{2p}{\bf C}_{2m}M^{2(p+m)-2}_{p}}{V_0 m^2}\left(\frac{\phi_{e}}{M_p}\right)^{2(p+m)-2}\right.\\ \left.~~~~~~~~~~~~
\displaystyle-\sum^{\infty}_{m=1}\sum^{\infty}_{n=1}\sum^{\infty}_{p=2}\frac{p{\bf C}_{2p}{\bf C}_{2m}{\bf C}_{2n}M^{2(p+m+n)-2}_{p}}{V^2_0 m^2}
\left(\frac{\phi_{e}}{M_p}\right)^{2(p+m+n)-2}\right]\approx \displaystyle \frac{V^{2}_{0}\Delta\phi}{4\sigma m^{2}M^{2}_{p}\phi_{e}},\end{array}\ee

where for both the cases convergence criteria of the series sum are imposed.

%%%%%%%%%%%%%%%%%%%%%%%%%%%%%%%%%%%%%%%%%%%%%%%%%%%%%%%%%%%%%%%%%%%%%%%%%%%%%%%%%%%%%%%%%%%%%%%%%%%%%%%%%%%%%%%%%%%%%%%%%%%%%%%%%%%%%%%%%%%%%%%%%%%%%%%%%%%%%%%%%%%%%%%%%%%%%%%%%%%%%%%%%%%
%%%%%%%%%%%%%%%%%%%%%%%%%%%%%%%%%%%%%%%%%%%%%%%%%%%%%%%%%%%%%%%%%%%%%%%%%%%%%%%%%%%%%%%%%%%%%%%%%%%%%%%%%%%%%%%%%%%%%%%%%%%%%%%%%%%%%%%%%%%%%%%%%%%%%%%%%%%%%%%%%%%%%%%%%%%


\begin{thebibliography}{99}
%\begin{references}

\bibitem{Green1}
  Superstring Theory. Vol. 1: Introduction - Green, Michael B. et al. Cambridge, Uk: Univ. Pr. ( 1987) 469 P. ( Cambridge Monographs On Mathematical Physics).

\bibitem{Green2} 
 Superstring Theory. Vol. 2: Loop Amplitudes,
 Anomalies And Phenomenology - Green, Michael B. et al. Cambridge, Uk: Univ. Pr. ( 1987) 596 P. ( Cambridge Monographs On Mathematical Physics).

\bibitem{Polchinski2}
String theory. Vol. 2: Superstring theory and beyond - Polchinski, J. Cambridge, UK: Univ. Pr. (1998) 531 p.

\bibitem{Maartens:2010ar}
  R.~Maartens and K.~Koyama,
  %``Brane-World Gravity,''
  Living Rev.\ Rel.\  {\bf 13} (2010) 5
  [arXiv:1004.3962 [hep-th]].
  %%CITATION = ARXIV:1004.3962;%%
  %146 citations counted in INSPIRE as of 28 Jun 2014

\bibitem{Brax:2004xh}
  P.~Brax, C.~van de Bruck and A.~-C.~Davis,
  %``Brane world cosmology,''
  Rept.\ Prog.\ Phys.\  {\bf 67} (2004) 2183
  [hep-th/0404011].
  %%CITATION = HEP-TH/0404011;%%
  %204 citations counted in INSPIRE as of 28 Jun 2014


\bibitem{Randall:1999vf}
  L.~Randall and R.~Sundrum,
  %``An Alternative to compactification,''
  Phys.\ Rev.\ Lett.\  {\bf 83} (1999) 4690
  [hep-th/9906064].
  %%CITATION = HEP-TH/9906064;%%
  %4899 citations counted in INSPIRE as of 20 Oct 2013itme


\bibitem{Randall:1999ee}
  L.~Randall and R.~Sundrum,
  %``A Large mass hierarchy from a small extra dimension,''
  Phys.\ Rev.\ Lett.\  {\bf 83} (1999) 3370
  [hep-ph/9905221].
  %%CITATION = HEP-PH/9905221;%%
  %5963 citations counted in INSPIRE as of 20 Oct 2013

\bibitem{Choudhury:2011sq}
  S.~Choudhury and S.~Pal,
  %``Brane inflation in background supergravity,''
  Phys.\ Rev.\ D {\bf 85} (2012) 043529
  [arXiv:1102.4206 [hep-th]].
  %%CITATION = ARXIV:1102.4206;%%
  %10 citations counted in INSPIRE as of 28 Jun 2014

\bibitem{Choudhury:2011rz}
  S.~Choudhury and S.~Pal,
  %``Reheating and leptogenesis in a SUGRA inspired brane inflation,''
  Nucl.\ Phys.\ B {\bf 857} (2012) 85
  [arXiv:1108.5676 [hep-ph]].
  %%CITATION = ARXIV:1108.5676;%%
  %8 citations counted in INSPIRE as of 28 Jun 2014 

\bibitem{Choudhury:2012ib}
  S.~Choudhury and S.~Pal,
  %``Brane inflation: A field theory approach in background supergravity,''
  J.\ Phys.\ Conf.\ Ser.\  {\bf 405} (2012) 012009
  [arXiv:1209.5883 [hep-th]].
  %%CITATION = ARXIV:1209.5883;%%
  %6 citations counted in INSPIRE as of 28 Jun 2014 

\bibitem{Choudhury:2013yg}
  S.~Choudhury and S.~Sengupta,
  %``Features of warped geometry in presence of Gauss-Bonnet coupling,''
  JHEP {\bf 1302} (2013) 136
  [arXiv:1301.0918 [hep-th]].
  %%CITATION = ARXIV:1301.0918;%%
  %9 citations counted in INSPIRE as of 29 Jun 2014 

\bibitem{Choudhury:2013aqa}
  S.~Choudhury and S.~SenGupta,
  %``A step towards exploring the features of Gravidilaton sector in string phenomenology via lightest Kaluza-Klein graviton mass,''
  arXiv:1311.0730 [hep-ph].
  %%CITATION = ARXIV:1311.0730;%%
  %1 citations counted in INSPIRE as of 29 Jun 2014 

\bibitem{Choudhury:2013eoa}
  S.~Choudhury, S.~Sadhukhan and S.~SenGupta,
  %``Collider constraints on Gauss-Bonnet coupling in warped geometry model,''
  arXiv:1308.1477 [hep-ph].
  %%CITATION = ARXIV:1308.1477;%%
  %3 citations counted in INSPIRE as of 29 Jun 2014


\bibitem{Choudhury:2014hna}
  S.~Choudhury, J.~Mitra and S.~SenGupta,
  %``Modulus stabilization in higher curvature dilaton gravity,''
 JHEP {\bf 1408} (2014) 004 [arXiv:1405.6826 [hep-th]].
  %%CITATION = ARXIV:1405.6826;%% 

\bibitem{Das:2013lqa}
  A.~Das and S.~SenGupta,
  %``126 GeV Higgs and ATLAS bound on the lightest graviton mass in Randall-Sundrum model,''
  arXiv:1303.2512 [hep-ph].
  %%CITATION = ARXIV:1303.2512;%%
  %4 citations counted in INSPIRE as of 29 Jun 2014 

\bibitem{Banerjee:2014wea}
  N.~Banerjee, S.~Lahiri and S.~SenGupta,
  %``Cosmology in multiply warped braneworld scenario,''
  Int.\ J.\ Mod.\ Phys.\ A {\bf 29} (2014) 13,  1450069.
  %%CITATION = IMPAE,A29,1450069;%% 

\bibitem{Minamitsuji:2005xs}
  M.~Minamitsuji, M.~Sasaki and D.~Langlois,
  %``Kaluza-Klein gravitons are negative energy dust in brane cosmology,''
  Phys.\ Rev.\ D {\bf 71} (2005) 084019
  [gr-qc/0501086].
  %%CITATION = GR-QC/0501086;%%
  %15 citations counted in INSPIRE as of 29 Jun 2014

\bibitem{Himemoto:2000nd}
  Y.~Himemoto and M.~Sasaki,
  %``Brane world inflation without inflaton on the brane,''
  Phys.\ Rev.\ D {\bf 63} (2001) 044015
  [gr-qc/0010035].
  %%CITATION = GR-QC/0010035;%%
  %103 citations counted in INSPIRE as of 29 Jun 2014

\bibitem{Mukhopadhyaya:2002jn}
  B.~Mukhopadhyaya, S.~Sen and S.~SenGupta,
  %``Does a Randall-Sundrum scenario create the illusion of a torsion free universe?,''
  Phys.\ Rev.\ Lett.\  {\bf 89} (2002) 121101
   [Erratum-ibid.\  {\bf 89} (2002) 259902]
  [hep-th/0204242].
  %%CITATION = HEP-TH/0204242;%%
  %47 citations counted in INSPIRE as of 01 Jul 2014 

\bibitem{Kar:2001eb}
  S.~Kar, P.~Majumdar, S.~SenGupta and S.~Sur,
  %``Cosmic optical activity from an inhomogeneous Kalb-Ramond field,''
  Class.\ Quant.\ Grav.\  {\bf 19} (2002) 677
  [hep-th/0109135].
  %%CITATION = HEP-TH/0109135;%%
  %38 citations counted in INSPIRE as of 01 Jul 2014 

\bibitem{Ghosh:2008vc}
  S.~Ghosh and S.~Kar,
  %``Bulk spacetimes for cosmological braneworlds with a time-dependent extra dimension,''
  Phys.\ Rev.\ D {\bf 80} (2009) 064024
  [arXiv:0812.1666 [gr-qc]].
  %%CITATION = ARXIV:0812.1666;%%
  %8 citations counted in INSPIRE as of 01 Jul 2014

\bibitem{Koley:2005nv}
  R.~Koley and S.~Kar,
  %``A Novel braneworld model with a bulk scalar field,''
  Phys.\ Lett.\ B {\bf 623} (2005) 244
   [Erratum-ibid.\ B {\bf 631} (2005) 199]
  [hep-th/0507277].
  %%CITATION = HEP-TH/0507277;%%
  %39 citations counted in INSPIRE as of 01 Jul 2014


\bibitem{Mukhanov:1981xt} 
  V.~F.~Mukhanov and G.~V.~Chibisov,
  %``Quantum Fluctuation and Nonsingular Universe. (In Russian),''
  JETP Lett.\  {\bf 33}, 532 (1981)
  [Pisma Zh.\ Eksp.\ Teor.\ Fiz.\  {\bf 33}, 549 (1981)].
  %%CITATION = JTPLA,33,532;%%
  %696 citations counted in INSPIRE as of 12 Mar 2014
   A.~A.~Starobinsky,
  %``Relict Gravitation Radiation Spectrum and Initial State of the Universe. (In Russian),''
  JETP Lett.\  {\bf 30}, 682 (1979)
  [Pisma Zh.\ Eksp.\ Teor.\ Fiz.\  {\bf 30}, 719 (1979)].
  %%CITATION = JTPLA,30,682;%%
  %780 citations counted in INSPIRE as of 12 Mar 2014

 \bibitem{Mukhanov:1990me} 
  V.~F.~Mukhanov, H.~A.~Feldman and R.~H.~Brandenberger,
  %``Theory of cosmological perturbations. Part 1. Classical perturbations. Part 2. Quantum theory of perturbations. Part 3. Extensions,''
  Phys.\ Rept.\  {\bf 215}, 203 (1992).
  %%CITATION = PRPLC,215,203;%%
  %1893 citations counted in INSPIRE as of 12 Mar 2014



\bibitem{Ade:2014xna} 
  P.~A.~R.~Ade {\it et al.}  [BICEP2 Collaboration],
  %``BICEP2 I: Detection Of B-mode Polarization at Degree Angular Scales,''
  arXiv:1403.3985 [astro-ph.CO].
  %%CITATION = ARXIV:1403.3985;%%


\bibitem{Liu:2014mpa}
  H.~Liu, P.~Mertsch and S.~Sarkar,
  %``Fingerprints of Galactic Loop I on the Cosmic Microwave Background,''
  Astrophys.\ J.\  {\bf 789} (2014) L29
  [arXiv:1404.1899 [astro-ph.CO]].
  %%CITATION = ARXIV:1404.1899;%%
  %34 citations counted in INSPIRE as of 04 Sep 2014 

\bibitem{Mortonson:2014bja}
  M.~J.~Mortonson and U.~Seljak,
  %``A joint analysis of Planck and BICEP2 B modes including dust polarization uncertainty,''
  arXiv:1405.5857 [astro-ph.CO].
  %%CITATION = ARXIV:1405.5857;%%
  %74 citations counted in INSPIRE as of 04 Sep 2014 

\bibitem{Flauger:2014qra}
  R.~Flauger, J.~C.~Hill and D.~N.~Spergel,
  %``Toward an Understanding of Foreground Emission in the BICEP2 Region,''
  JCAP {\bf 1408} (2014) 039
  [arXiv:1405.7351 [astro-ph.CO]]. 

\bibitem{Adam:2014gaa}
  R.~Adam {\it et al.}  [ Planck Collaboration],
  %``Planck intermediate results. XXXII. The relative orientation between the magnetic field and structures traced by interstellar dust,''
  arXiv:1409.6728 [astro-ph.GA].

\bibitem{Ade:2015tva}
  P.~A.~R.~Ade {\it et al.}  [BICEP2 and Planck Collaborations],
  %``A Joint Analysis of BICEP2/Keck Array and Planck Data,''
  %Submitted to: Phys.Rev.Lett.
  arXiv:1502.00612 [astro-ph.CO].


\bibitem{Ade:2015lrj}
  P.~A.~R.~Ade {\it et al.}  [Planck Collaboration],
  %``Planck 2015. XX. Constraints on inflation,''
  arXiv:1502.02114 [astro-ph.CO].
  %%CITATION = ARXIV:1502.02114;%%
  %10 citations counted in INSPIRE as of 21 Feb 2015

\bibitem{Lyth:1996im}
  D.~H.~Lyth,
  %``What would we learn by detecting a gravitational wave signal in the cosmic microwave background anisotropy?,''
  Phys.\ Rev.\ Lett.\  {\bf 78} (1997) 1861
  [hep-ph/9606387].
  %%CITATION = HEP-PH/9606387;%%
  %354 citations counted in INSPIRE as of 28 Jun 2014

\bibitem{Baumann:2009ds}
  D.~Baumann,
  %``TASI Lectures on Inflation,''
  arXiv:0907.5424 [hep-th].
  %%CITATION = ARXIV:0907.5424;%%
  %174 citations counted in INSPIRE as of 29 Jun 2014 


\bibitem{Baumann:2014nda}
  D.~Baumann and L.~McAllister,
  %``Inflation and String Theory,''
  arXiv:1404.2601 [hep-th].
  %%CITATION = ARXIV:1404.2601;%%
  %28 citations counted in INSPIRE as of 29 Jun 2014 

\bibitem{Linde:2014nna}
  A.~Linde,
  %``Inflationary Cosmology after Planck 2013,''
  arXiv:1402.0526 [hep-th].
  %%CITATION = ARXIV:1402.0526;%%
  %35 citations counted in INSPIRE as of 05 Aug 2014 

\bibitem{Kallosh:2014xwa}
  R.~Kallosh, A.~Linde and A.~Westphal,
  %``Chaotic Inflation in Supergravity after Planck and BICEP2,''
  arXiv:1405.0270 [hep-th].
  %%CITATION = ARXIV:1405.0270;%%
  %18 citations counted in INSPIRE as of 05 Aug 2014


\bibitem{Kallosh:2014rga}
  R.~Kallosh, A.~Linde and D.~Roest,
  %``Large Field Inflation and Double $\alpha$-Attractors,''
  arXiv:1405.3646 [hep-th].
  %%CITATION = ARXIV:1405.3646;%%
  %5 citations counted in INSPIRE as of 05 Aug 2014 


\bibitem{Lyth:1998xn}
  D.~H.~Lyth and A.~Riotto,
  %``Particle physics models of inflation and the cosmological density perturbation,''
  Phys.\ Rept.\  {\bf 314} (1999) 1
  [hep-ph/9807278].
  %%CITATION = HEP-PH/9807278;%%
  %1223 citations counted in INSPIRE as of 05 Aug 2014

\bibitem{Mazumdar:2010sa}
  A.~Mazumdar and J.~Rocher,
  %``Particle physics models of inflation and curvaton scenarios,''
  Phys.\ Rept.\  {\bf 497} (2011) 85
  [arXiv:1001.0993 [hep-ph]].
  %%CITATION = ARXIV:1001.0993;%%
  %181 citations counted in INSPIRE as of 05 Aug 2014


\bibitem{Martin:2014vha}
  J.~Martin, C.~Ringeval and V.~Vennin,
  %``Encyclopædia Inflationaris,''
  Phys.\ Dark Univ.\  (2014)
  [arXiv:1303.3787 [astro-ph.CO]].
  %%CITATION = ARXIV:1303.3787;%%
  %93 citations counted in INSPIRE as of 05 Aug 2014 

\bibitem{Martin:2013nzq}
  J.~Martin, C.~Ringeval, R.~Trotta and V.~Vennin,
  %``The Best Inflationary Models After Planck,''
  JCAP {\bf 1403} (2014) 039
  [arXiv:1312.3529 [astro-ph.CO]].
  %%CITATION = ARXIV:1312.3529;%%
  %27 citations counted in INSPIRE as of 05 Aug 2014

\bibitem{Chialva:2014rla}
  D.~Chialva and A.~Mazumdar,
  %``Super-Planckian excursions of the inflaton and quantum corrections,''
  arXiv:1405.0513 [hep-th].
  %%CITATION = ARXIV:1405.0513;%%
  %9 citations counted in INSPIRE as of 05 Aug 2014 

\bibitem{Biswas:2011ar}
  T.~Biswas, E.~Gerwick, T.~Koivisto and A.~Mazumdar,
  %``Towards singularity and ghost free theories of gravity,''
  Phys.\ Rev.\ Lett.\  {\bf 108} (2012) 031101
  [arXiv:1110.5249 [gr-qc]].
  %%CITATION = ARXIV:1110.5249;%%
  %95 citations counted in INSPIRE as of 05 Aug 2014

\bibitem{Biswas:2013cha}
  T.~Biswas, A.~Conroy, A.~S.~Koshelev and A.~Mazumdar,
  %``Generalized ghost-free quadratic curvature gravity,''
  Class.\ Quant.\ Grav.\  {\bf 31} (2014) 015022
   [Erratum-ibid.\  {\bf 31} (2014) 159501]
  [arXiv:1308.2319 [hep-th]].
  %%CITATION = ARXIV:1308.2319;%%
  %9 citations counted in INSPIRE as of 05 Aug 2014 


\bibitem{Assassi:2013gxa}
  V.~Assassi, D.~Baumann, D.~Green and L.~McAllister,
  %``Planck-Suppressed Operators,''
  arXiv:1304.5226 [hep-th].
  %%CITATION = ARXIV:1304.5226;%%
  %13 citations counted in INSPIRE as of 29 Jun 2014


\bibitem{Choudhury:2013iaa}
  S.~Choudhury and A.~Mazumdar,
  %``An accurate bound on tensor-to-scalar ratio and the scale of inflation,''
  Nucl.\ Phys.\ B {\bf 882} (2014) 386
  [arXiv:1306.4496 [hep-ph]].
  %%CITATION = ARXIV:1306.4496;%%
  %25 citations counted in INSPIRE as of 28 Jun 2014 

\bibitem{Choudhury:2014kma}
  S.~Choudhury and A.~Mazumdar,
  %``Reconstructing inflationary potential from BICEP2 and running of tensor modes,''
  arXiv:1403.5549 [hep-th].
  %%CITATION = ARXIV:1403.5549;%%
  %40 citations counted in INSPIRE as of 28 Jun 2014

\bibitem{Choudhury:2014wsa}
  S.~Choudhury and A.~Mazumdar,
  %``Sub-Planckian inflation & large tensor to scalar ratio with $r\geq 0.1$,''
  arXiv:1404.3398 [hep-th].
  %%CITATION = ARXIV:1404.3398;%%
  %9 citations counted in INSPIRE as of 28 Jun 2014 

\bibitem{Choudhury:2013woa}
  S.~Choudhury and A.~Mazumdar,
  %``Primordial blackholes and gravitational waves for an inflection-point model of inflation,''
  Phys.\ Lett.\ B {\bf 733} (2014) 270
  [arXiv:1307.5119 [astro-ph.CO]].
  %%CITATION = ARXIV:1307.5119;%%
  %4 citations counted in INSPIRE as of 28 Jun 2014

\bibitem{Choudhury:2014hua}
  S.~Choudhury,
  %``Inflamagnetogenesis redux: Unzipping sub-Planckian inflation via various cosmoparticle probes,''
  Phys.\ Lett.\ B {\bf 735} (2014) 138
  [arXiv:1403.0676 [hep-th]].
  %%CITATION = ARXIV:1403.0676;%% 

\bibitem{Choudhury:2013jya}
  S.~Choudhury, A.~Mazumdar and S.~Pal,
  %``Low & High scale MSSM inflation, gravitational waves and constraints from Planck,''
  JCAP {\bf 1307} (2013) 041
  [arXiv:1305.6398 [hep-ph]].
  %%CITATION = ARXIV:1305.6398;%%
  %21 citations counted in INSPIRE as of 29 Jun 2014

\bibitem{Choudhury:2014sxa}
  S.~Choudhury, A.~Mazumdar and E.~Pukartas,
  %``Constraining ${\cal N}=1$ supergravity inflationary framework with non-minimal Kähler operators,''
  JHEP {\bf 1404} (2014) 077
  [arXiv:1402.1227 [hep-th]].
  %%CITATION = ARXIV:1402.1227;%%
  %6 citations counted in INSPIRE as of 29 Jun 2014 

\bibitem{Choudhury:2014uxa}
  S.~Choudhury,
  %``Constraining ${\cal N}=1$ supergravity inflation with non-minimal K\"ahler operators using $\delta N$ formalism,''
  JHEP {\bf 04} (2014) 105
  [arXiv:1402.1251 [hep-th]].
  %%CITATION = ARXIV:1402.1251;%%
  %5 citations counted in INSPIRE as of 29 Jun 2014

\bibitem{Choudhury:2011jt}
  S.~Choudhury and S.~Pal,
  %``Fourth level MSSM inflation from new flat directions,''
  JCAP {\bf 1204} (2012) 018
  [arXiv:1111.3441 [hep-ph]].
  %%CITATION = ARXIV:1111.3441;%%
  %3 citations counted in INSPIRE as of 29 Jun 2014

\bibitem{Cheung:2007st}
  C.~Cheung, P.~Creminelli, A.~L.~Fitzpatrick, J.~Kaplan and L.~Senatore,
  %``The Effective Field Theory of Inflation,''
  JHEP {\bf 0803} (2008) 014
  [arXiv:0709.0293 [hep-th]].
  %%CITATION = ARXIV:0709.0293;%%
  %292 citations counted in INSPIRE as of 29 Jun 2014 

\bibitem{Weinberg:2008hq}
  S.~Weinberg,
  %``Effective Field Theory for Inflation,''
  Phys.\ Rev.\ D {\bf 77} (2008) 123541
  [arXiv:0804.4291 [hep-th]].
  %%CITATION = ARXIV:0804.4291;%%
  %145 citations counted in INSPIRE as of 29 Jun 2014

\bibitem{Tsujikawa:2014mba}
  S.~Tsujikawa,
  %``The effective field theory of inflation/dark energy and the Horndeski theory,''
  arXiv:1404.2684 [gr-qc].
  %%CITATION = ARXIV:1404.2684;%%
  %2 citations counted in INSPIRE as of 29 Jun 2014 

\bibitem{Senatore:2010wk}
  L.~Senatore and M.~Zaldarriaga,
  %``The Effective Field Theory of Multifield Inflation,''
  JHEP {\bf 1204} (2012) 024
  [arXiv:1009.2093 [hep-th]].
  %%CITATION = ARXIV:1009.2093;%%
  %76 citations counted in INSPIRE as of 29 Jun 2014 

\bibitem{LopezNacir:2011kk}
  D.~Lopez Nacir, R.~A.~Porto, L.~Senatore and M.~Zaldarriaga,
  %``Dissipative effects in the Effective Field Theory of Inflation,''
  JHEP {\bf 1201} (2012) 075
  [arXiv:1109.4192 [hep-th]].
  %%CITATION = ARXIV:1109.4192;%%
  %37 citations counted in INSPIRE as of 29 Jun 2014 

\bibitem{Agarwal:2013rva}
  N.~Agarwal, R.~H.~Ribeiro and R.~Holman,
  %``Why does the effective field theory of inflation work?,''
  JCAP {\bf 1406} (2014) 016
  [arXiv:1311.0869 [hep-th]].
  %%CITATION = ARXIV:1311.0869;%%
  %2 citations counted in INSPIRE as of 29 Jun 20 

\bibitem{Baumann:2011nm}
  D.~Baumann and D.~Green,
  %``Supergravity for Effective Theories,''
  JHEP {\bf 1203} (2012) 001
  [arXiv:1109.0293 [hep-th]].
  %%CITATION = ARXIV:1109.0293;%%
  %18 citations counted in INSPIRE as of 29 Jun 2014 

\bibitem{Creminelli:2013xfa}
  P.~Creminelli, R.~Emami, M.~Simonović and G.~Trevisan,
  %``ISO(4,1) Symmetry in the EFT of Inflation,''
  JCAP {\bf 1307} (2013) 037
  [arXiv:1304.4238 [hep-th]].
  %%CITATION = ARXIV:1304.4238;%%
  %2 citations counted in INSPIRE as of 29 Jun 2014 

\bibitem{Khosravi:2012qg}
  N.~Khosravi,
  %``Effective Field Theory of Multi-Field Inflation a la Weinberg,''
  JCAP {\bf 1205} (2012) 018
  [arXiv:1203.2266 [hep-th]].
  %%CITATION = ARXIV:1203.2266;%%
  %8 citations counted in INSPIRE as of 12 Jul 2014

\bibitem{Liddle:1998jc}
  A.~R.~Liddle, A.~Mazumdar and F.~E.~Schunck,
  %``Assisted inflation,''
  Phys.\ Rev.\ D {\bf 58} (1998) 061301
  [astro-ph/9804177].
  %%CITATION = ASTRO-PH/9804177;%%
  %317 citations counted in INSPIRE as of 05 Aug 2014 

\bibitem{Copeland:1999cs}
  E.~J.~Copeland, A.~Mazumdar and N.~J.~Nunes,
  %``Generalized assisted inflation,''
  Phys.\ Rev.\ D {\bf 60} (1999) 083506
  [astro-ph/9904309].
  %%CITATION = ASTRO-PH/9904309;%%
  %123 citations counted in INSPIRE as of 05 Aug 2014 

\bibitem{Kanti:1999ie}
  P.~Kanti and K.~A.~Olive,
  %``Assisted chaotic inflation in higher dimensional theories,''
  Phys.\ Lett.\ B {\bf 464} (1999) 192
  [hep-ph/9906331].
  %%CITATION = HEP-PH/9906331;%%
  %104 citations counted in INSPIRE as of 05 Aug 2014 

\bibitem{Kanti:1999vt}
  P.~Kanti and K.~A.~Olive,
  %``On the realization of assisted inflation,''
  Phys.\ Rev.\ D {\bf 60} (1999) 043502
  [hep-ph/9903524].
  %%CITATION = HEP-PH/9903524;%%
  %121 citations counted in INSPIRE as of 05 Aug 2014 

\bibitem{Mazumdar:2001mm}
  A.~Mazumdar, S.~Panda and A.~Perez-Lorenzana,
  %``Assisted inflation via tachyon condensation,''
  Nucl.\ Phys.\ B {\bf 614} (2001) 101
  [hep-ph/0107058].
  %%CITATION = HEP-PH/0107058;%%
  %199 citations counted in INSPIRE as of 05 Aug 2014 

\bibitem{Green:1999vv}
  A.~M.~Green and J.~E.~Lidsey,
  %``Generalized compactification and assisted dynamics of multiscalar field cosmologies,''
  Phys.\ Rev.\ D {\bf 61} (2000) 067301
  [astro-ph/9907223].
  %%CITATION = ASTRO-PH/9907223;%%
  %42 citations counted in INSPIRE as of 05 Aug 2014 

\bibitem{Malik:1998gy}
  K.~A.~Malik and D.~Wands,
  %``Dynamics of assisted inflation,''
  Phys.\ Rev.\ D {\bf 59} (1999) 123501
  [astro-ph/9812204].
  %%CITATION = ASTRO-PH/9812204;%%
  %95 citations counted in INSPIRE as of 05 Aug 2014



\bibitem{Dimopoulos:2005ac}
  S.~Dimopoulos, S.~Kachru, J.~McGreevy and J.~G.~Wacker,
  %``N-flation,''
  JCAP {\bf 0808} (2008) 003
  [hep-th/0507205].
  %%CITATION = HEP-TH/0507205;%%
  %302 citations counted in INSPIRE as of 05 Aug 2014

\bibitem{Cicoli:2014sva}
  M.~Cicoli, K.~Dutta and A.~Maharana,
  %``N-flation with Hierarchically Light Axions in String Compactifications,''
  arXiv:1401.2579 [hep-th].
  %%CITATION = ARXIV:1401.2579;%%
  %16 citations counted in INSPIRE as of 05 Aug 2014 

\bibitem{Easther:2005zr}
  R.~Easther and L.~McAllister,
  %``Random matrices and the spectrum of N-flation,''
  JCAP {\bf 0605} (2006) 018
  [hep-th/0512102].
  %%CITATION = HEP-TH/0512102;%%
  %110 citations counted in INSPIRE as of 05 Aug 2014

\bibitem{Stelle:1976gc}
  K.~S.~Stelle,
  %``Renormalization of Higher Derivative Quantum Gravity,''
  Phys.\ Rev.\ D {\bf 16} (1977) 953.
  %%CITATION = PHRVA,D16,953;%%
  %986 citations counted in INSPIRE as of 05 Aug 2014 

\bibitem{Stelle:1977ry}
  K.~S.~Stelle,
  %``Classical Gravity with Higher Derivatives,''
  Gen.\ Rel.\ Grav.\  {\bf 9} (1978) 353.
  %%CITATION = GRGVA,9,353;%%
  %453 citations counted in INSPIRE as of 05 Aug 2014 

\bibitem{Nunez:2004ts}
  A.~Nunez and S.~Solganik,
  %``Ghost constraints on modified gravity,''
  Phys.\ Lett.\ B {\bf 608} (2005) 189
  [hep-th/0411102].
  %%CITATION = HEP-TH/0411102;%%
  %62 citations counted in INSPIRE as of 05 Aug 2014

\bibitem{Brax:2005jv}
  P.~Brax and J.~Martin,
  %``Shift symmetry and inflation in supergravity,''
  Phys.\ Rev.\ D {\bf 72} (2005) 023518
  [hep-th/0504168].
  %%CITATION = HEP-TH/0504168;%%
  %32 citations counted in INSPIRE as of 05 Aug 2014 

\bibitem{Choudhury:2013zna}
  S.~Choudhury, T.~Chakraborty and S.~Pal,
  %``Higgs inflation from new Kähler potential,''
  Nucl.\ Phys.\ B {\bf 880} (2014) 155
  [arXiv:1305.0981 [hep-th]].
  %%CITATION = ARXIV:1305.0981;%%
  %4 citations counted in INSPIRE as of 05 Aug 2014



\bibitem{Shiromizu:1999wj}
  T.~Shiromizu, K.~-i.~Maeda and M.~Sasaki,
  %``The Einstein equation on the 3-brane world,''
  Phys.\ Rev.\ D {\bf 62} (2000) 024012
  [gr-qc/9910076].
  %%CITATION = GR-QC/9910076;%%
  %1112 citations counted in INSPIRE as of 29 Jun 2014

\bibitem{Allahverdi:2006iq}
  R.~Allahverdi, K.~Enqvist, J.~Garcia-Bellido and A.~Mazumdar,
  %``Gauge invariant MSSM inflaton,''
  Phys.\ Rev.\ Lett.\  {\bf 97} (2006) 191304
  [hep-ph/0605035].
  %%CITATION = HEP-PH/0605035;%%
  %210 citations counted in INSPIRE as of 05 Aug 2014

\bibitem{Allahverdi:2006we}
  R.~Allahverdi, K.~Enqvist, J.~Garcia-Bellido, A.~Jokinen and A.~Mazumdar,
  %``MSSM flat direction inflation: Slow roll, stability, fine tunning and reheating,''
  JCAP {\bf 0706} (2007) 019
  [hep-ph/0610134].
  %%CITATION = HEP-PH/0610134;%%
  %155 citations counted in INSPIRE as of 05 Aug 2014

 \bibitem{Burgess:2005sb} 
  C.~P.~Burgess, et.al.
  %``Multiple inflation, cosmic string networks and the string landscape,''
  JHEP {\bf 0505}, 067 (2005)
  [hep-th/0501125].
  %%CITATION = HEP-TH/0501125;%%
  %77 citations counted in INSPIRE as of 20 Mar 2014

  \bibitem{Planck-1}
  P.~A.~R.~Ade {\it et al.}  [Planck Collaboration],
  %``Planck 2013 results. XVI. Cosmological parameters,''
  arXiv:1303.5076 [astro-ph.CO].
  %%CITATION = ARXIV:1303.5076;%%
  %106 citations counted in INSPIRE as of 01 May 2013
  
  \bibitem{Planck-infl}
   P.~A.~R.~Ade {\it et al.}  [Planck Collaboration],
  %``Planck 2013 results. XXII. Constraints on inflation,''
  arXiv:1303.5082 [astro-ph.CO].
  %%CITATION = ARXIV:1303.5082;%%
  %27 citations counted in INSPIRE as of 22 Apr 2013hva

 \bibitem{Bramante:2014rva}
  J.~Bramante, S.~Downes, L.~Lehman and A.~Martin,
  %``Last stand of single small field inflation,''
  Phys.\ Rev.\ D {\bf 90} (2014) 2,  023530
  [arXiv:1405.7563 [astro-ph.CO]].
  %%CITATION = ARXIV:1405.7563;%%
  %6 citations counted in INSPIRE as of 21 Feb 2015

\bibitem{Antusch:2014cpa}
  S.~Antusch and D.~Nolde,
  %``BICEP2 implications for single-field slow-roll inflation revisited,''
  JCAP {\bf 1405} (2014) 035
  [arXiv:1404.1821 [hep-ph]].
  %%CITATION = ARXIV:1404.1821;%%
  %42 citations counted in INSPIRE as of 23 Feb 2015

  \bibitem{Hotchkiss:2011gz}
  S.~Hotchkiss, A.~Mazumdar and S.~Nadathur,
  %``Observable gravitational waves from inflation with small field excursions,''
  JCAP {\bf 1202} (2012) 008
  [arXiv:1110.5389 [astro-ph.CO]].
  %%CITATION = ARXIV:1110.5389;%%
  %64 citations counted in INSPIRE as of 22 Feb 2015



%\end{references}

\end{thebibliography}
\end{document}